\documentclass[twoside]{scrartcl}
\usepackage[utf8]{inputenc}
\usepackage{units}
\usepackage{array}
\usepackage{graphicx}
\usepackage{epstopdf}
\usepackage{float}
\usepackage{amsmath}
\usepackage{amssymb}
\usepackage{esint}
\usepackage{multicol}
\usepackage{changepage}

\usepackage{hyperref}
\usepackage{xcolor}
\usepackage{float}
\usepackage{placeins}

\usepackage{listings}

\usepackage{hyphenat}

\usepackage{lineno}

\usepackage[justification=justified]{caption}
\usepackage{prettyref}
\newrefformat{sfig}{Supp. Fig. \ref{#1}}
\newrefformat{tab}{Supp. Table \ref{#1}}
\DeclareCaptionLabelFormat{supptable}{#1 S#2} 
\DeclareCaptionLabelFormat{suppfig}{#1 S#2}

\newcommand{\trace}{\mathrm{tr}}
\newcommand{\dev}{\mathrm{dev}}

\newcommand{\unittensor}{\mathrm{I}}

\newcommand{\maxima}{{\texttt{GNU-Maxima} }}

\newcommand{\la}{\ell_a}
\newcommand{\lb}{\ell_b}

\newcommand{\epsM}{\epsilon}
\newcommand{\epsa}{\epsM_a}
\newcommand{\epsb}{\epsM_b}

\newcommand{\LX}{X}
\newcommand{\LY}{Y}

\newcommand{\epsx}{\epsM_x}
\newcommand{\epsy}{\epsM_y}
\newcommand{\epsz}{\epsM_z}

\newcommand{\lo}{\ell_0}

\newcommand{\Etis}{E}

\newcommand{\Gtis}{G}
\newcommand{\DGtis}{\Delta\Gtis}
\newcommand{\Ktis}{K}
\newcommand{\DKtis}{\Delta\Ktis}
\newcommand{\nutis}{N}

\newcommand{\so}{\sigma_0}

\newcommand{\Mucortex}{\mu}

\newcommand{\pdd}{{p^\mathrm{2D}}}

\newcommand{\Kcytodd}{{k_\mathrm{c}}}

\newcommand{\LZ}{Z}
\newcommand{\eface}{\epsilon^\mathrm{facet}}

\newcommand{\ptd}{{p^\mathrm{3D}}}
\newcommand{\ptdz}{{p^\mathrm{3D}_0}}

\newcommand{\Edd}{e}

\newcommand{\Eddh}{e_H}
\newcommand{\nudd}{\nu_\mathrm{c}}

\newcommand{\etadd}{\widehat{\eta}}
\newcommand{\cbetadd}{\widehat{c}}
\newcommand{\betadd}{\widehat{\beta}}
\newcommand{\kdd}{\widehat{k}}

\newcommand{\hor}{\mathrm{H}}
\newcommand{\lat}{\mathrm{L}}

\newcommand{\gh}{{g_\hor}}
\newcommand{\kh}{{k_\hor}}
\newcommand{\soh}{\sigma_{0\hor}}

\newcommand{\gl}{{g_\lat}}
\newcommand{\kl}{{k_\lat}}
\newcommand{\sol}{\sigma_{0\lat}}

\newcommand{\nonaffinity}{\mathcal{N}}

\newcommand{\red}[1]{\textcolor{black}{{#1}}}
\newcommand{\blue}[1]{\textcolor{black}{{#1}}}

\makeatletter

\pdfpageheight\paperheight
\pdfpagewidth\paperwidth

\makeatother

\begin{document}

\begin{flushleft}
{\Large
\textbf{Mapping cell cortex rheology to tissue rheology, and vice-versa
}
}
\newline
\\
\'Etienne Moisdon\textsuperscript{1,2},
Pierre Seez\textsuperscript{1,2},
Camille No\^us\textsuperscript{2},
Fran\c{c}ois Molino\textsuperscript{3,2},
Philippe Marcq\textsuperscript{4,2}, 
Cyprien~Gay\textsuperscript{1,2,*}
\\
\bigskip
\textbf{1}
Laboratoire Mati\`ere et Syst\`emes Complexes, UMR 7057, CNRS and Universit\'e Paris Cité, 75205 Paris cedex 13, France
\\ 
\textbf{2}
Laboratoire Cogitamus, Paris, France   
\\ 
\textbf{3}
Laboratoire Charles Coulomb, UMR 5221, CNRS and Université de Montpellier, Place Eugène Bataillon, F-34095 Montpellier, France
\\
\textbf{4}
PMMH, CNRS, ESPCI Paris, PSL University, Sorbonne Université, Université Paris Cité, F-75005 Paris, France
\\ 
\textbf{*}
Author for correspondence: \texttt{cyprien.gay@univ-paris-diderot.fr}
\end{flushleft}

\date{\today}

\setcounter{section}{0}
\setcounter{figure}{0}
\setcounter{table}{0}
\setcounter{equation}{0}


\section*{Abstract}


The mechanics of biological tissues
mainly proceeds from the cell cortex rheology.
A direct, explicit link between cortex rheology and tissue rheology 
remains lacking, yet would be instrumental in understanding how 
modulations of cortical mechanics may impact tissue 
mechanical behaviour.
Using an ordered geometry built on 3D hexagonal, incompressible cells,
we build a mapping relating
the cortical rheology to the monolayer tissue rheology.
Our approach shows that the tissue 
low frequency elastic modulus
is proportional to the rest tension of the cortex,
as expected from the physics of liquid foams
\red{as well as of tensegrity structures}.
\red{A fractional visco-contractile cortex rheology
is predicted to yield a high-frequency fractional visco-elastic 
monolayer rheology, where such a fractional behaviour has been 
recently observed experimentally at each scale separately.}
In particular cases, the mapping may be inverted, allowing to
derive from a given tissue rheology the underlying cortex rheology. 
Interestingly, applying the same approach to a 2D hexagonal tiling fails,
which suggests that the 2D character of planar cell cortex-based 
models may be unsuitable to account for realistic monolayer rheologies.
We provide quantitative predictions, amenable to 
experimental tests through standard perturbation assays of
cortex constituents, 
and hope to foster new, challenging mechanical
experiments on cell monolayers.

\section{Introduction}

The mechanical determinants of embryonic development
have received considerable attention in recent years
\cite{Heisenberg2013,Goodwin2021,Valet2021}, 
with an emphasis on ingredients such as surface tension, 
fluid flows, active stresses or boundary conditions.
Of note, the complex rheology of tissues \cite{Verdier2009} 
harbours immediate consequences for morphogenetic processes 
\cite{Petridou2019}, as the response of cells to forces 
within the tissue depends on, \textit{e.g.}, whether the tissue rheology 
is elastic rather than viscous on the relevant time scale.

Since \textit{in vivo} measurements remain arduous, rheologists have naturally turned
to \textit{in vitro} tissues such as epithelial cell aggregates 
and monolayers, for relative ease of use and control.
In the case of cellular aggregates, a variety of techniques has been brought to bear, among which 
micropipette aspiration~\cite{Guevorkian2010prl},
parallel plate compression~\cite{stirbat2013} or magnetic rheometry~\cite{Mary2022}, unraveling a complex rheological
behaviour that can be viewed as
combinations of elastic, viscous, plastic and fractional elements.
Epithelial cell monolayers cultured on a flat substrate have revealed an active~\cite{Vincent2015,Nier2018} and
viscoelastic~\cite{Tlili2020} rheology on a time scale longer than one hour.
In the absence of a substrate,
suspended epithelial monolayers held by adhesive
micromanipulators~\cite{Harris2013,Khalilgharibi2019} 
have been characterized by a  composite fractional model~\cite{Bonfanti2020} on a time scale shorter than
a few minutes.

Tissues being assemblies of mechanically coupled cells,
one generally expects tissue rheology to depend on cell rheology~\cite{Verdier2009}.
In turn, cell rheology is highly dependent on cytoskeletal rheology~\cite{Pullarkat2007},
both known since early micro-rheological measurements~\red{\cite{Fabry2001,Lenormand2004}} \cite{Desprat2005,Balland2006}
to display a power-law behaviour. 
The cell cortex, principally made of actin, myosin,
and their cross-linkers~\cite{Chugh2018}, is generally thought 
to behave as a viscoelastic material,
liquid at time scales large compared to the turnover times of its constituents~\cite{Salbreux2012}.
Frequency-dependent measurements in single cells submitted to  uniaxial compression have confirmed that the cell cortex behaves as a viscoelastic liquid~\cite{FF2016}.
Similar experiments have shown that the cell cortex Poisson ratio  ranges typically
from $0.2$ to $0.6$, decreasing with frequency~\cite{Mokbel2020}.
Within tissues, cellular cortices in contact 
form cell-cell junctions.
Using optical tweezers, the viscoelastic time
of a cell junction has been measured in the \textit{Drosophila melanogaster} embryo,
and is of the order of one minute \cite{Clement2017}.
More recently, the out-of-plane rheology of the cell-cell junction in a cell doublet has been
studied experimentally thanks to the introduction of novel micromanipulation techniques~\cite{Esfahani2021}.
The mechanical response of apical cortices has been probed by atomic force microscopy in epithelial cell monolayers \cite{Pietuch2013},
and by laser ablation in \textit{Caenorhabditis elegans} and zebrafish \cite{Saha2016}.

\red{Because}
most perturbations of tissue mechanical behavior operate 
in practice at the sub-cellular level,
\red{it is essential (and} 
it remains a challenge\red{)} 
to relate the mechanical properties and descriptions at the microscopic (cell) and macroscopic (tissue) levels.
The elastic properties of inert cellular materials 
can be computed as a function 
of cell elasticity~\cite{Gibson2014} 
in the case of solid walls
or as a function of cell surface tensions in the case of liquid foams~\cite{Weaire01,Cantat2013}.
The study of living cellular materials remains less advanced\red{,
although, \textit{e.g.}, the mechanics of lungs has been investigated 
in the framework of hexagonal networks of elastic springs
\cite{Mead1970,Stamenovic1990,Cavalcante2005}.}
A popular cell-based computational model of a cell monolayer
is the cell vertex model~\cite{Fletcher2014,Alt2017}.
This model is based on an energy function,
together with a viscous friction on the substrate,
which eases computation.
The latter is an external force.
Hence, \red{in this} model 
\red{the monolayer}
by itself is conservative
and the corresponding
effective macroscopic rheology is purely elastic.
The corresponding elastic moduli
are expected to scale like the rest tension of the cell cortex divided by the cell size,
by analogy with a classical result for liquid foams~\cite{Princen1986,Reynelt1993}.
The tissue-scale elastic stress based on the 2D vertex model has been computed considering
in-plane~\cite{Ishihara2017} and out-of-plane deformations~\cite{Murisic2015}.
To the best of our knowledge, the case of the 3D vertex model~\cite{Okuda2015} has not been addressed from this perspective.
When topological transitions such as cell intercalations 
(also called cell rearrangements) 
are allowed, the rheological behavior of a 2D vertex model 
becomes fluid~\cite{Tong2021}, %
as anticipated from a theoretical viewpoint~\cite{Tlili2015}
(see also \cite{Ishihara2017,Grossman2021} for predictions of nonlinear 
rheological behaviour when topological transitions are included).
Tissue behaviours \red{such as} 
tissue deformations and possibly cell intercalations~\cite{Rauzi2008}
may then be reproduced 
using different and/or slowly varying tensions among 
the various cell-cell junctions, 
thus possibly mimicking internal gene expression levels in cells.
However, vertex models do not consider 
\red{non-trivial} 
cortical rheologies
such as observed in studies of single cells or single junctions.
As argued in~\cite{Lenne2021},
models of cell junction mechanics should now include dissipative effects.

In the present work, 
we take into account the truly complex mechanical
behaviour of the cellular cortex and use it in a cell-based
model to derive the corresponding rheological behaviour at the scale of
the tissue. 
Defining a regular cell network geometry
allows to derive this connection analytically, leading to general
expressions mapping the cortex to the tissue rheology.
Existing results are discussed in view of these expressions.
As we address the in-plane rheology of the cell monolayer, viewed as a
2D material, our results connect naturally 
to suspended monolayer experiments~\cite{Harris2013,Khalilgharibi2019}.

As studies focusing on measurements of tissue or cell response to
short time stimuli 
remain infrequent~\cite{Harris2013,Khalilgharibi2019,Clement2017,Chanet2017}, 
we focus on timescales up to 10 minutes,
well below the viscoelastic time of about 1 hour,
characteristic of epithelial cell monolayers~\cite{Tlili2020}.
Beyond that timescale, the tissue becomes fluid,
as a result of such mechanisms
as cell rearrangement, cell division and apoptosis~\cite{Ranft2010}.
We expect~\cite{Tlili2015} that 
this fluid response expected at long time scales
will combine in series 
with the solid viscoelastic
behaviour established in the present work.

The article is organized as follows.
We first describe the (three-dimensional) geometry,
the assumed cell properties 
and the calculation methodology (Section~\ref{sec:model}).
We then present the resulting monolayer viscoelastic properties
and compare them with recent measurements (Section~\ref{sec:mapping}).
The inverse mapping from the monolayer properties
to the cell-scale properties is possible in various ways
using some additional assumptions
(Section~\ref{sec:inverse:mapping}).
We finally conclude and discuss our results
(Section~\ref{sec:discussion:perspectives}).

\section{Model}
\label{sec:model}

Our simple approach relies on several ingredients,
summarized on Fig.~\ref{fig:ingredients} 
and detailed in Section~\ref{sec:ingredients}.
Notations are defined on
Figs.~\ref{fig:ingredients}-\ref{fig:monolayer:top:view}
and in their captions.
The method for deriving the resulting monolayer moduli
is then exposed in Section~\ref{sec:resolution}.

\subsection{Ingredients}
\label{sec:ingredients}

\subsubsection*{Geometry}
\begin{figure}[t]
\textbf{a.}
\includegraphics[scale=0.45]{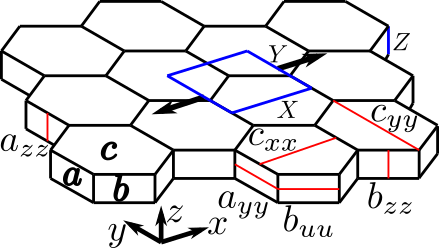}
\hspace*{1cm}
\textbf{b.}
\includegraphics[scale=1.3]{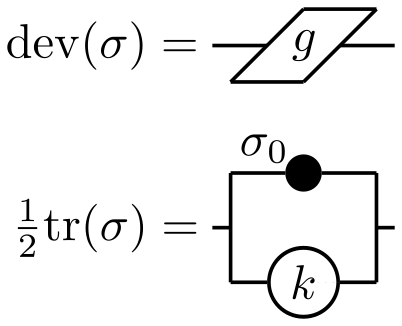}
\caption{\textbf{Model ingredients: a.} Geometry: hexagonal cells tile the plane, with 
\red{flat}
facets $a$, $b$ and $c$.
A representative, rectangular region of the monolayer is drawn in blue, with a volume $V = X Y Z$.
Relevant facets are labeled $a$, $b$ (lateral) and $c$
(horizontal).
Principal axes of the facets are drawn in red.
\textbf{b.} Rheological diagrams for the cortex
corresponding to Eq.~\eqref{eq:rheo:cortex:2D:dev}
and~\eqref{eq:rheo:cortex:2D:tr},
with the traceless and isotropic components of the stress.
Here, as well as in Fig.~\ref{fig:monolayer:rheological:diagrams},
the (complex) shear modulus $g$ is 
represented by a parallelogram, 
the (complex) compression modulus $k$ by a circle,
the rest tension $\so$ by a filled disk. 
}
\label{fig:ingredients}
\end{figure}

The cell monolayer is assumed to be a regular arrangement of hexagonal 
cells with constant height $Z$, see Fig.~\ref{fig:ingredients},
with lateral facets $a$ and $b$
\red{
(often called \textit{cell-cell junctions}),
} 
and horizontal facets $c$
(both basal and apical).

Keeping in mind  the boundary conditions imposed onto the suspended monolayer
studied in \cite{Harris2013,Khalilgharibi2019,Bonfanti2020},
we assume that the monolayer is being stretched or compressed
uniformly along the $x$-axis with no stress applied along the $y$-axis
or along the vertical direction, $z$. In this model, all cells 
therefore remain identical at all times.
Possible extensions of these boundary conditions
are briefly discussed in Section~\ref{sec:discussion:perspectives}.

\subsubsection*{Intra-cellular material}

We consider the intra-cellular material to behave as an inviscid, incompressible fluid, with pressure $\ptd$
and constant volume $V$, expressed in terms of the dimensions of 
the representative region, depicted on Fig.~\ref{fig:ingredients}a,     as:
\begin{equation}
V=\LX\, \LY\, \LZ \,.
\label{eq:volume:X:Y:Z}
\end{equation}
Although cell volume is known to fluctuate in MDCK monolayers over a 
timescale of $10^3$~s \cite{Zehnder2015,Zehnder2015a}, this effect is 
neglected over the shorter time scales considered in this work.

\subsubsection*{Cortex tension}

The cell cortex is known to spontaneously develop a tension,
which typically stabilizes at some value
that depends on the amount of myosin and ATP
present and on the network architecture~\cite{Chugh2018}. 
Based on this fact, we choose to include a fixed, 
2D-isotropic tension,
denoted $\so$, in the isotropic part, $\mathrm{tr}(\sigma)$, of
the cortical stress (see Fig.~\ref{fig:ingredients}b).
A typical value of cortex tension is
$\so = 0.3 \, \mathrm{mN \, m^{-1}}$~\cite{Salbreux2012}. 

Since this tension may differ between horizontal ($\soh$)
and lateral ($\sol$) facets, we introduce the notation
\begin{equation}
\Psi \equiv \frac{\soh}{\sol} \,.
\label{eq:def:Psi}
\end{equation}
The tension aspect ratio $\Psi$ is known to affect the tissue behaviour
and may trigger monolayer folding~\cite{Sui2018,Messal2019}.

\subsubsection*{Cortex rheology}

In this work, we consider the cortex material to be 2D-isotropic. As a consequence, 
any variation of the in-plane stress in the cortex, $\sigma$,
around its rest value, $\so$,
in response to an in-plane deformation, $\eface$,
can be modeled, to linear order,
by a pair of (frequency dependent) complex moduli~\red{\cite{Wikipedia2022elastenglish,landau1986theory}}.
We choose the compression modulus, $k$\red{~\cite{Wkp2022elasticbulkmodulus}},
and the shear modulus, $g$\red{~\cite{Wkp2022elasticshearmodulus}}. 
\red{The variation of the in-plane stress tensor $\sigma$ of the cortex
around its rest value
is then classically~\cite{Wkp2022hookeslaw,landau1986theory}
expressed (here in 2D) as:}
\begin{eqnarray}
\red{\sigma -\sigma_0\,\unittensor} &\red{=}& \red{2g\,\eface +(k-g)\,\trace(\eface)\,\unittensor \,,}
\label{eq:rheo:cortex:2D:k:g}
\\
&\red{=}& \red{2g\,\dev(\eface) + k\,\trace(\eface)\,\unittensor \,,}
\label{eq:rheo:cortex:2D:k:g:dev}
\end{eqnarray}
\red{where $\unittensor$ is the 2D identity tensor
and where $\eface$ is decomposed into 
its isotropic part $\frac{1}{2} \trace(\eface)\, \red{\unittensor}$
and its traceless part $\dev(\eface) \equiv \eface -\frac{1}{2} \trace(\eface)\, \red{\unittensor}$.}
\red{Similarly, the traceless and isotropic parts of $\sigma$} 
\red{are given by:}
\begin{eqnarray}
\dev(\sigma) &=& 2g\,\dev(\eface) \,,
\label{eq:rheo:cortex:2D:dev}
\\
\red{\frac12}\,\trace(\sigma) &=& \so + k\,\trace(\eface) \,,
\label{eq:rheo:cortex:2D:tr}
\end{eqnarray}
\red{which is depicted schematically on Fig.~\ref{fig:ingredients}b.}
\red{Let us recall another classical pair of coefficients, 
namely the Young modulus, $\Edd$,
and the Poisson ratio, $\nudd$.
They are best suited for a situation in which the cortex is being deformed in one of its in-plane directions, 
while no force is exerted in the other in-plane direction.
The Young modulus is then the ratio of the stress to the deformation in the active direction,
while the Poisson ratio is the relative amount of induced deformation in the other direction. Note that for most materials, the sign of the latter is the opposite of that of the former, and the Poisson ratio is conventionally positive for a material with such a behaviour.
More precisely, the variation of the in-plane stress tensor $\sigma$ of the cortex
around its rest value
is expressed in terms of $\Edd$ and $\nudd$ as:}
\begin{eqnarray}
\red{\sigma -\sigma_0\,\unittensor} &\red{=}& \red{\frac{\Edd}{1+\nudd}\,\eface +\frac{\Edd\,\nudd}{1-\nudd^2}\,\trace(\eface)\,\unittensor \,,}
\label{eq:rheo:cortex:2D:e:nu}
\\
&\red{=}& \red{\frac{\Edd}{1+\nudd}\,\dev(\eface) + \frac{\Edd}{2\,(1-\nudd)}\,\trace(\eface)\,\unittensor \,.}
\label{eq:rheo:cortex:2D:e:nu:dev}
\end{eqnarray}
They are related to $k$ and $g$ through the following classical expressions, valid in two
dimensions~\red{\cite{Wikipedia2022elastenglish,landau1986theory}}: 
\begin{eqnarray}
\frac{1}{\Edd} &=& \frac{1}{4k} + \frac{1}{4g} \,,
\\
\nudd &=& \frac{k-g}{k+g} \,.
\label{eq:nudd:k:g}
\end{eqnarray}

\blue{In the incompressible cortex limit ($\nudd\rightarrow 1$),
which is probably rather unrealistic
according to the work by Mokbel \textit{et al.}~\cite{Mokbel2020},
Eqs.~(\ref{eq:rheo:cortex:2D:e:nu}-\ref{eq:rheo:cortex:2D:e:nu:dev})
behave in the classical manner~\cite{landau1986theory}:
the denominator in the last term diverges
and is compensated by a vanishing isotropic component of the deformation (trace),
to accomodate any finite isotropic component of the stress on the left-hand side.
} 

Due to the inner dissipative components of the cortex,
it is reasonable to expect that the amplitudes of its moduli
increase with frequency (see Appendix~\ref{sec:frequency:dependence}
for a discussion of this point).
In what follows, in order for the tension $\sigma$ 
to remain equal to $\sigma_0$ at rest
\red{within the framework of Eqs.~(\ref{eq:rheo:cortex:2D:dev}-\ref{eq:rheo:cortex:2D:tr})},
we shall assume more precisely
that the amplitudes of cortex shear
\red{($g$)}
and compression 
\red{($k$)}
moduli vanish in the low frequency limit $\omega \to 0$.
%

\subsubsection*{Rest state}

As a consequence of the above assumptions,
when the monolayer is left at rest,
\textit{i.e.}, with no applied in-plane stress,
hexagons are fully symmetric,
and elementary force balance considerations
lead to the following relations between
the cell pressure ($\ptdz$),
the dimensions of the representative volume
($\LX_0$, $\LY_0$ and $\LZ = \LZ_0$),
the tension aspect ratio $\Psi$,
and the cell volume ($V = V_0$) at rest:%

\begin{eqnarray}
\LX_0 &=& \sqrt{3}\,\lo \,,
\label{eq:X0}
\\
\LY_0 &=& \frac32\,\lo \,,
\label{eq:Y0}
\\
\frac{\LZ_0}{\LX_0} &=& \Psi \,,
\label{eq:aspect:ratio:sol:soh}
\\
V &=& \frac{3\sqrt{3}}{2}\,\lo^2\LZ_0
= \frac{9\,\Psi}{2}\,\lo^3 \,,
\label{eq:cell:volume}
\\
\ptdz &=& \frac{4\,\sol}{\sqrt{3}\,\lo}
= \frac{4\,\soh}{3\,\LZ_0} \,,
\label{eq:pressure:rest:sol:soh}
\end{eqnarray}
where $\lo$ is the rest value
of both $\la$ and $\lb$ (see Fig.~\ref{fig:monolayer:top:view}).

Importantly, Eq.~\eqref{eq:aspect:ratio:sol:soh} shows 
that the rest tension ratio $\Psi$
coincides with the cell aspect ratio $\LZ_0/\LX_0$ at rest.
Larger values of $\Psi$ thus correspond to a columnar epithelium,
smaller values to a squamous epithelium, while the monolayer
will be made of cuboidal cells when $\Psi \sim 1$.

\subsection{Model resolution}
\label{sec:resolution}

\subsubsection*{The linear response is isotropic}

In the present work, the monolayer is viewed
as a two-dimensional object in the $xy$-plane,
with in-plane stress components integrated over its thickness $Z$.
With the present choice of cortex 2D isotropy
and cytoplasm 2D (and even 3D) isotropy,
the only possible source of monolayer 2D anisotropy
could lie in the spatial arrangement of cells:
here, the regular honeycomb geometry
warrants the 2D isotropy of the monolayer linear mechanical response
(see \cite{Wilson1972,Gibson2014,Boal2012}
for a derivation in the case of an elastic solid).

\begin{figure}[th!]
\textbf{a.}
\includegraphics[scale=0.45]{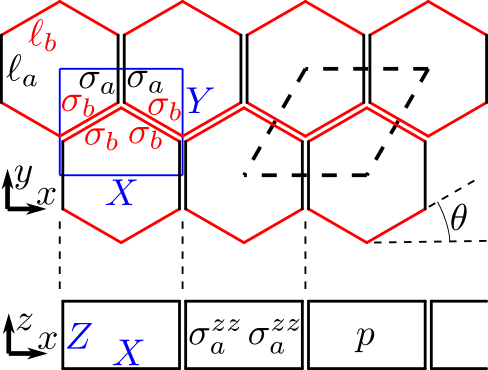}
\hspace*{1cm}
\textbf{b.}
\includegraphics[scale=0.45]{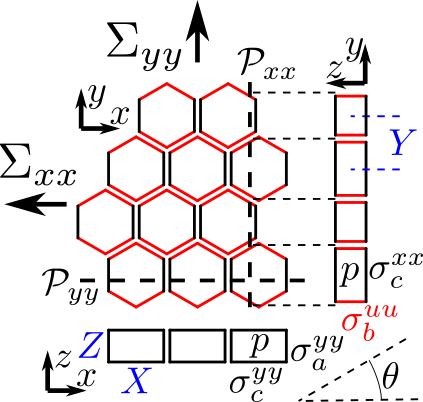}
\caption{\textbf{Views of the monolayer, force balance and macroscopic stress.}\newline\textbf{a.}
\textbf{Top:} top view. The facet lengths $\la$ and $\lb$
and the orientation $\theta$ of $b$ facets, are shown.
The repeating unit is shown as a dashed parallelogram.
A simpler representative rectangular region,
of same surface area $\LX\LY$, is shown in blue
as in Fig.~\ref{fig:ingredients}.
Force balance between three neighbouring cells:
at the meeting point within the blue rectangle,
all six cortex tensions add up vectorially to zero,
which yields Eq.~\eqref{eq:force:balance:six:cortex:tensions}
(notations $\sigma_a^{yy}$ and $\sigma_b^{uu}$ have been simplified for clarity).
\textbf{Bottom:} vertical cross-section along in the $xz$-plane.
The force balance in the vertical direction
implies that the pressure $p$ in the cytoplasm
integrated over the whole cell width $X$
balances the tensions $\sigma_a^{zz}$ 
on either side of the cell,
as expressed by Eq.~\eqref{eq:pressure:sigmazz:lateral}.\\
\textbf{b.} Macroscopic stress.
The $yy$ component of the in-plane macroscopic stress, $\Sigma_{yy}$,
can be expressed by counting all forces 
that are exerted across a perpendicular plane $\mathcal{P}_{yy}$, represented as a dashed horizontal line.
Such forces are readily enumerated from the monolayer cross-section 
in the $xz$ plane
shown below the top view.
They include the pressure $p$ in the cell
(integrated over the monolayer thickness, $\LZ$),
the tension $\sigma_c^{yy}$ of both horizontal layers,
as well as the tensions $\sigma_a^{yy}$ 
of the relevant lateral facets,
integrated over the facet height $\LZ$.
This provides the expression of Eq.~\eqref{eq:stress:macro:yy}.
The expression of the $xx$ component $\Sigma_{xx}$,
provided by Eq.~\eqref{eq:stress:macro:xx},
is obtained in a similar way
(see monolayer cross-section 
in the $yz$ plane
on the right-hand side),
except that the lateral facet tensions, $\sigma_b^{uu}$, 
must be projected onto the $x$~axis (angle $\theta$, see also Fig.~\ref{fig:ingredients}).
}
\label{fig:monolayer:top:view}
\end{figure}

\subsubsection*{Force balances}

Velocities and accelerations in such monolayers are so small
that inertial contributions to force balances are negligible
(a situation similar to low Reynolds number regimes in fluids).
As for in-plane forces, the balance between all six cortex tensions
that are exerted on a given vertical edge,
such as the vertex at the center of the blue rectangle 
in the top-view of Fig.~\ref{fig:ingredients}a
or~\ref{fig:monolayer:top:view}a, can be expressed as:
\begin{equation}
2 \sigma_{a}^{yy}\sin(\theta_{a})- 4 \sigma_{b}^{uu}\sin(\theta_{b})=0 \,,
\label{eq:force:balance:six:cortex:tensions}
\end{equation}
where $\sigma_{a}^{yy}$ and $\sigma_{b}^{uu}$
are the horizontal components of the tensions
in facets $a$ and $b$,
see Fig.~\ref{fig:monolayer:top:view} for details.

In the present, simplified geometry, all cells play the same role
and all adjacent cells have equal pressures.
Lateral cortices,
\red{being under tension,} 
are correspondingly flat.
Because the monolayer is assumed to be subjected
to zero external forces in the $z$-direction,
the force balance at mid-height implies
that the pressure $\ptd$ in the cell
is compensated by the 
$z$-tension in the lateral facets at all times
(in-plane surface area $\LX\LY$, horizontal lengths $\la$ and $\lb$):
\begin{equation}
\ptd\,\LX\LY = 2\,\sigma_{a}^{zz}\,\la +4\,\sigma_{b}^{zz}\,\lb \,.
\label{eq:pressure:sigmazz:lateral}
\end{equation}
In the context of the cross-section depicted 
in the bottom of Fig.~\ref{fig:monolayer:top:view}a,
this simply reads: $\ptd\,X = 2\,\sigma_{a}^{zz}$.

For the sake of simplicity, we here consider
that horizontal (apical and basal) facets
remain flat despite the larger pressure in the cell
as compared to the outside medium,
as if these facets were not flexible.

\subsubsection*{Deformation of the rectangular representative volume}

As mentioned in Section~\ref{sec:ingredients},
the monolayer is being stretched along the $x$-axis,
with no forces applied along the perpendicular directions.
As a result, the distances $\LX=2\,\lb\,\cos\theta$ 
and $\LY=\la +\lb\,\sin\theta$
depicted on Fig.~\ref{fig:monolayer:top:view}a
(with initial values $\LX_0$ and $\LY_0$, respectively)
are related to the monolayer deformation components 
along the $x$- and $y$-axes through:
\begin{eqnarray}
\epsx&=&\frac{\LX-{\LX}_{0}}{{\LX}_{0}} \,,
\label{eq:epsx}
\\
\epsy&=&\frac{\LY-{\LY}_{0}}{{\LY}_0} \,.
\label{eq:epsy}
\end{eqnarray}

In our calculation, we assume that the horizontal facets 
(index $c$) deform homogeneously and affinely,
that is, in the same proportions as the monolayer,
although that is not strictly true
\red{(see Appendix~\ref{sec:affinity} for detail)}.
As a result, the in-plane stress in the horizontal facets
can be expressed from
Eqs.~(\ref{eq:rheo:cortex:2D:dev}-\ref{eq:rheo:cortex:2D:tr})
and Eqs.~(\ref{eq:epsx}-\ref{eq:epsy})
in terms of their compression
\red{($\kh$)}
and shear 
\red{($\gh$)}
moduli as:
\begin{eqnarray}
\sigma_c^{xx} &=& \soh + \kh\,(\epsx+\epsy) +\gh\,(\epsx-\epsy) \,,
\label{eq:rheo:cortex:c:xx}
\\
\sigma_c^{yy} &=& \soh + \kh\,(\epsx+\epsy) +\gh\,(\epsy-\epsx) \,.
\label{eq:rheo:cortex:c:yy}
\end{eqnarray}

For the lateral facets (indices $a$ and $b$), we similarly define:
\begin{eqnarray}
\epsa&=&\frac{\la-\lo}{\lo} \,,
\label{eq:epsa}
\\
\epsb&=&\frac{\lb-\lo}{\lo} \,,
\label{eq:epsb}
\\
\epsz&=&\frac{\LZ-\LZ_0}{\LZ_0} \,.
\label{eq:epsz}
\end{eqnarray}
Again, using Eqs.~(\ref{eq:rheo:cortex:2D:dev}-\ref{eq:rheo:cortex:2D:tr})
and Eqs.~(\ref{eq:epsa},\ref{eq:epsb},\ref{eq:epsz}),
we express the in-plane stress components of the lateral facets
in terms of their compression \red{($\kl$)}
and shear 
\red{($\gl$)}
moduli as:
\begin{eqnarray}
\sigma_a^{yy} &=& \sol + \kl\,(\epsa+\epsz) +\gl\,(\epsa-\epsz) \,,
\label{eq:rheo:cortex:a:yy}
\\
\sigma_a^{zz} &=& \sol + \kl\,(\epsz+\epsa) +\gl\,(\epsz-\epsa)  \,,
\label{eq:rheo:cortex:a:zz}
\\
\sigma_b^{uu} &=& \sol + \kl\,(\epsb+\epsz) +\gl\,(\epsb-\epsz) \,,
\label{eq:rheo:cortex:b:uu}
\\
\sigma_b^{zz} &=& \sol + \kl\,(\epsz+\epsb) +\gl\,(\epsz-\epsb) \,.
\label{eq:rheo:cortex:b:zz}
\end{eqnarray}

\subsubsection*{Expression of monolayer in-plane stress}

The hexagonal symmetry of the monolayer
and our choice to exert forces along the $x$-axis
ensure that the macroscopic stress $\Sigma$ 
has principal components along axes $x$ and $y$.
The stress along the $x$-axis is readily expressed
by considering all forces that cut a section 
perpendicular to the $x$-axis,
as illustrated on Fig.~\ref{fig:monolayer:top:view}b:
\begin{equation}
\Sigma_{xx} = 2\sigma_{c}^{xx} -\ptd\,\LZ
+\frac{4\lb\LZ}{\LX\LY}\, \sigma_{b}^{uu}\,\cos^2\theta  \,.
\label{eq:stress:macro:xx}
\end{equation}
Similarly for the stress along the $y$-axis:
\begin{equation}
\Sigma_{yy} = 2\sigma_{c}^{yy} -\ptd\,\LZ
+\frac{2\la\LZ}{\LX\LY}\, \sigma_{a}^{yy}
+\frac{4\lb\LZ}{\LX\LY}\, 
\sigma_{b}^{uu}\,\sin^2\theta  \,.
\label{eq:stress:macro:yy}
\end{equation}

\subsubsection*{Monolayer in-plane complex moduli}

Combining Eqs.~(\ref{eq:volume:X:Y:Z}-\ref{eq:rheo:cortex:b:zz})
and imposing that $\Sigma_{yy}$ remains equal to zero
when the monolayer is deformed in the $x$-direction,
we derive as a function of $\epsx$ the expressions 
of the in-plane stress $\Sigma_{xx}$ in the $x$-direction
and of the deformation $\epsy$ in the $y$-direction.
From these two quantities $\Sigma_{xx}$ and $\epsy$,
we derive the complex Young modulus $\Etis$ and the complex
Poisson ratio $\nutis$
of the monolayer, viewed as a two-dimensional material:
\begin{eqnarray}
\Etis &=& \frac{\Sigma_{xx}}{\epsx} \,,
\\
\nutis &=& -\frac{\epsy}{\epsx} \,.
\end{eqnarray}
By combining the Poisson ratio and the Young modulus,
we may then derive any other complex modulus,
for instance the in-plane compression and shear moduli of the monolayer:
\begin{eqnarray}
\Ktis &=& \frac{\Etis}{2\,(1-\nutis)} \,,
\\
\Gtis &=& \frac{\Etis}{2\,(1+\nutis)} \,.
\end{eqnarray}

\section{Mapping cortex 
(micro-)rheology to tissue (macro-)rheology}
\label{sec:mapping}

\subsection{General result}
\label{sec:K:G:from:k:g:so:H:L}

In practice, the calculations 
\red{
outlined in Section~\ref{sec:resolution}
} 
were performed with \texttt{GNU-Maxima},
see Appendix~\ref{sec:approche:maxima}.

\begin{figure}[th!]
\begin{center}
\textbf{a.}\quad
\includegraphics[width=0.3\columnwidth]{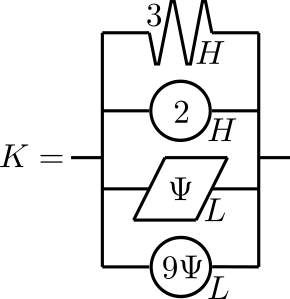}
\qquad
\textbf{b.}\quad
\includegraphics[width=0.35\columnwidth]{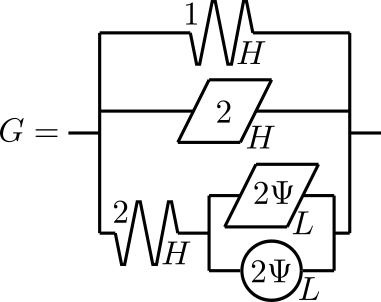}
\medskip
\medskip
\\
\textbf{c.}\quad
\includegraphics[width=0.35\columnwidth]{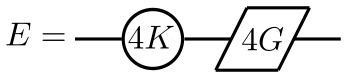}
\end{center}
\caption{\textbf{Rheological diagrams for the monolayer}.
\textbf{a.} and \textbf{b.} The monolayer compression modulus $\Ktis$ and shear modulus $\Gtis$
can be viewed as a mechanical arrangement of cortex properties.
The symbols are those of Fig.~\ref{fig:ingredients}.
Note that cortical tensions $\so$ in Fig.~\ref{fig:ingredients} 
become springs in the present figure
(with identical physical units).
The coefficients are those of Eqs.~\eqref{eq:K:s0:g:k:L:H} 
and~\eqref{eq:G:s0:g:k:L:H}, respectively.
$\Psi$ is defined by 
Eq.~\eqref{eq:def:Psi}.
Indices $H$ and $L$ refer to horizontal \textit{vs.} lateral facets.
\textbf{c.} The monolayer Young modulus $\Etis$
can be viewed as equivalent to $4$ times its compression modulus $\Ktis$ in series with $4$ times its shear modulus $\Gtis$. 
}
\label{fig:monolayer:rheological:diagrams}
\end{figure}

We now express the \red{$2$D} tissue moduli
as obtained following the path outlined above.
\red{
For the sake of clarity, 
we express $\Etis$ and $\nutis$
in terms of $\Ktis$ and $\Gtis$
which have simpler expressions:
} 
\begin{eqnarray}
\Ktis &=& 3\soh +2\kh 
+\Psi\,\kl +9\Psi\,\gl \,,
\label{eq:K:s0:g:k:L:H}
\\
\Gtis &=& \soh+2\gh +\frac{1}
{\frac{1}{2 \soh} +\frac{1}{2\Psi\,\kl +2\Psi\,\gl}} \,,
\label{eq:G:s0:g:k:L:H}
\\
\frac{1}{\Etis} &=& \frac{1}{4\,\Ktis} + \frac{1}{4\,\Gtis} \,,
\label{eq:E:s0:g:k:L:H}
\\
\nutis &=& \frac{\Ktis-\Gtis}{\Ktis+\Gtis} \,.
\label{eq:nu:s0:g:k:L:H}
\end{eqnarray}
where indices $H$ and $L$ 
\red{for rest} tensions \red{($\so$)}, 
\red{compression ($k$) and shear ($g$)} moduli
refer to horizontal and lateral facets, respectively,
see Eqs.~\eqref{eq:def:Psi},
(\ref{eq:rheo:cortex:c:xx},
\ref{eq:rheo:cortex:c:yy})
and (\ref{eq:rheo:cortex:a:yy}-\ref{eq:rheo:cortex:b:zz})
for reference.
Expressions~(\ref{eq:K:s0:g:k:L:H}-\ref{eq:E:s0:g:k:L:H}) 
can be represented conveniently as rheological diagrams, see Fig.~\ref{fig:monolayer:rheological:diagrams}. They map cortex 
(micro-) rheology to tissue (macro-) rheology. 
\red{In the absence of tissue prestress
\blue{as assumed for the rest state characterized 
by Eqs.~(\ref{eq:X0}-\ref{eq:pressure:rest:sol:soh})}, 
the 3D moduli $K$, $G$ and $E$ may be obtained from the above 
2D moduli by dividing by the monolayer thickness.}

\blue{The monolayer compression modulus $\Ktis$ 
depends obviously on the compression modulus $\kh$ of the horizontal (in-plane) facets.
} 
Somewhat surprisingly, 
\blue{it also}
depends on the shear modulus of lateral facets, $\gl$.
That reflects the fact that elongating the monolayer in an isotropic manner 
causes cells to flatten, 
hence lateral facets are elongated horizontally and \red{shrink} vertically,
\red{
see Fig.~\ref{fig:def:modes} in Appendix~\ref{sec:affinity}.
} 
\blue{In fact, even the lateral facet \textit{surface area}
is then required to change
due to the constant volume assumption, expressed by Eq.~\eqref{eq:cell:volume}.
That explains that the monolayer compression modulus $K$
also depends on the lateral facet compression modulus $\kl$.
As a result, in the limit of incompressible horizontal or lateral cortices
($\kh\rightarrow\infty$, $\kl\rightarrow\infty$,
which is probably not realistic~\cite{Mokbel2020}),
the monolayer would become incompressible 
($\Ktis\rightarrow\infty$, $\nutis\rightarrow 1$).
} 

As stated above (see Eqs.~(\ref{eq:def:Psi},\ref{eq:aspect:ratio:sol:soh})), 
the cell aspect ratio reflects the ratio $\Psi$ 
of the spontaneous tensions of horizontal and lateral facets.
Remarkably, while $\Ktis$ is dominated by the moduli of lateral facets
for large values of $\Psi$ (columnar cells)
and by the moduli of horizontal facets
for small values of $\Psi$ (flat cells),
by contrast $\Gtis$ is dominated
by the moduli of horizontal facets
for both monolayers made of very columnar 
and very flat cells ;
only for monolayers made of cuboidal cells
does $\Gtis$ depend substantially on the moduli of lateral facets.

\subsection{Low frequency limit}

In the low frequency limit, since $\gl$, $\kl$, $\gh$ and $\kh$ tend to zero,
tissue moduli are predicted to depend only on the tension of horizontal facets,
$\soh$, and to be real numbers:
\begin{eqnarray}
\Ktis_0 &=& 3\,\soh  \,,
\label{eq:Ktis:low:freq}
\\
\Gtis_0 &=& \soh  \,,
\label{eq:Gtis:low:freq}
\\
\Etis_0 &=& 3 \, \soh  \,,
\label{eq:Etis:low:freq}
\\
\nutis_0 &=& 1/2  \,,
\label{eq:nutis:low:freq}
\end{eqnarray}
The monolayer behaves as a purely elastic material in this limit.

In addition, we find that 
the above in-plane, 2D moduli are proportional 
to the cortex rest tension
$\soh$, as expected
from the physics of liquid bubble monolayers
at time scales where bubbles and films are stable~\cite{Weaire01,Cantat2013}.
This is of course expected from dimensional arguments,
as both the cortex and the monolayer are two-dimensional 
and their moduli correspondingly 
have the same physical units (N/m).
Remarkably, however, tissue moduli do not even depend on
the cell aspect ratio $\Psi$,
\red{
in other words on the horizontal to lateral facet tension ratio.
} 

\red{An additional interpretation of these results 
is that cell monolayers as well as liquid foams 
are examples of \textit{tensegrity structures}~\cite{Ingber2014}
which contain prestress, even at rest.
In this context, in the case of elastic structures,
it has been known for several decades
that the linear elastic moduli
are proportional to the prestress~\cite{Wang2002}.
In the context of liquid foams or cell monolayers,
some elements are under finite compression (bubble gas, cell cytoplasm)
while other elements are correspondingly under finite tension (soap films, cell cortices).
Note, however, that the origin of the prestress differs.
In elastic tensegrity structures, it results from the slight mismatch 
between each constitutive element individual rest size
and its rest size within the structure.
In other words, it results from some built-in geometrical frustration.
In liquid foams and cell monolayers, by constrast,
the constitutive elements 
\blue{under tension (soap films, cell cortices)}
are fluid,
\blue{\textit{i.e.,}}
have no finite individual rest size.
The prestress 
\blue{in the structure} 
results from the contractile nature of these particular fluids:
surface tension for liquid films, and myosin activity for cell 
\blue{cortices}. 
In other words, liquid foams and cell monolayers are examples 
of what we may call \textit{contractile fluid tensegrity structures}.
} 

Two aspects of the above prediction (\ref{eq:Ktis:low:freq}-\ref{eq:nutis:low:freq}) are testable.\\
\textit{(i)}. 
If at least two deformation modes
are accessible to assess the monolayer behaviour,
then the Poisson ratio can be derived,
or equivalently the following ratios can be tested:
\begin{equation}
\Ktis_0 = \Etis_0 = 3 \, \Gtis_0  \,.
\label{eq:low:freq:Ktis:Etis:Gtis:ratios}
\end{equation}
\textit{(ii)}. 
As blebbistatin (or other inhibitors of contractility)
lower the rest tension of the acto-myosin cortex~\cite{Chugh2018},
then the predicted proportionality of the monolayer elastic moduli
to the cell cortex rest tension
can be tested by applying such drugs (see, \textit{e.g.},
\cite{Vincent2015}, for a study of the effect of contractility
inhibition on epithelial cell monolayer tension).

In physiological conditions,
on timescales beyond about 10~minutes, tissues
are known to display biological phenomena not considered
in the present model, such as cell rearrangements, cell divisions
and cell extrusions.
Correspondingly, we shall choose $10^{-2}$ Hz
as the lower frequency bound in graphs.

\subsection{High frequency limit}

\begin{figure}[ht!]
\centering
\textbf{a.}
\includegraphics[scale=0.4]{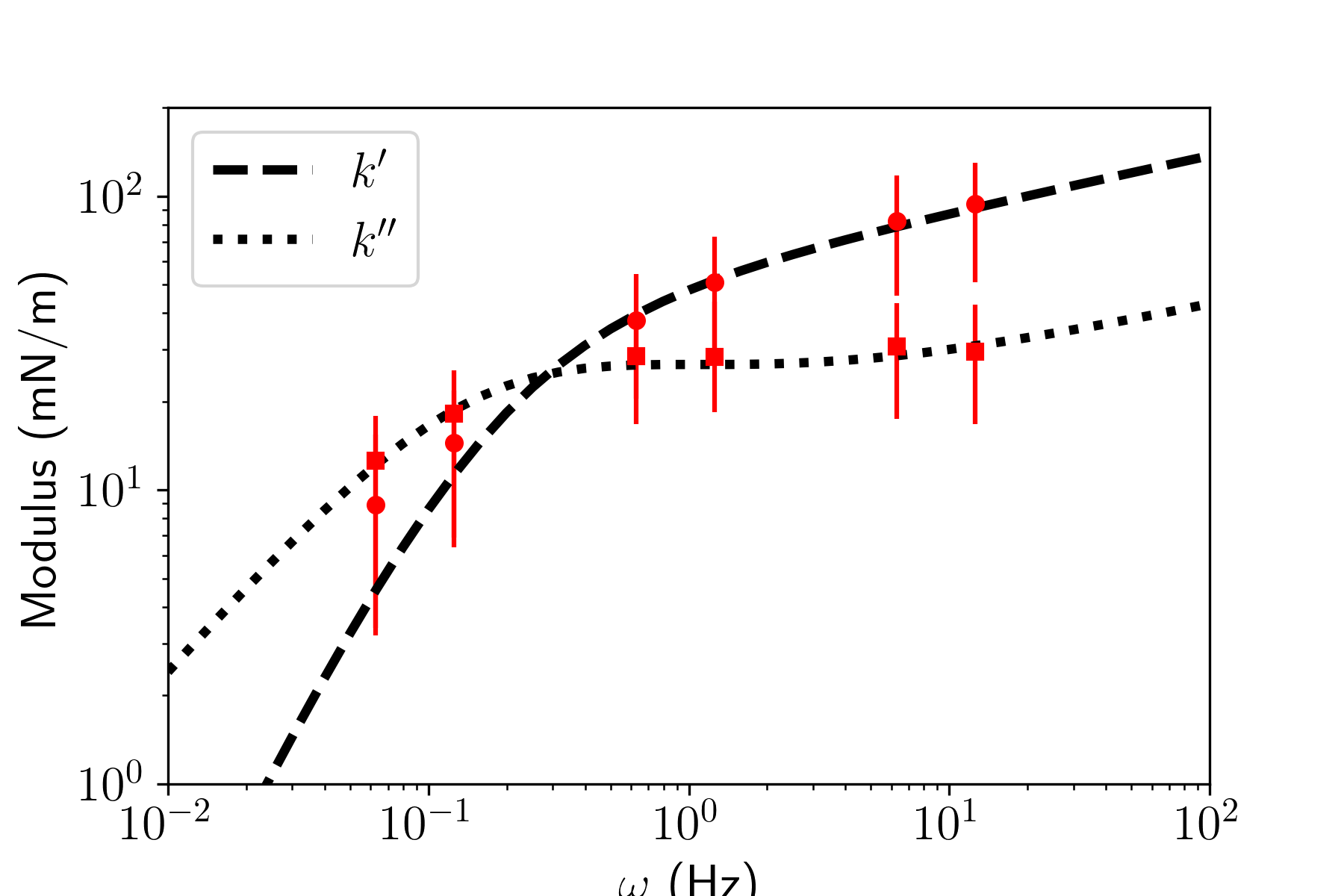}
\textbf{b.}
\includegraphics[scale=0.4]{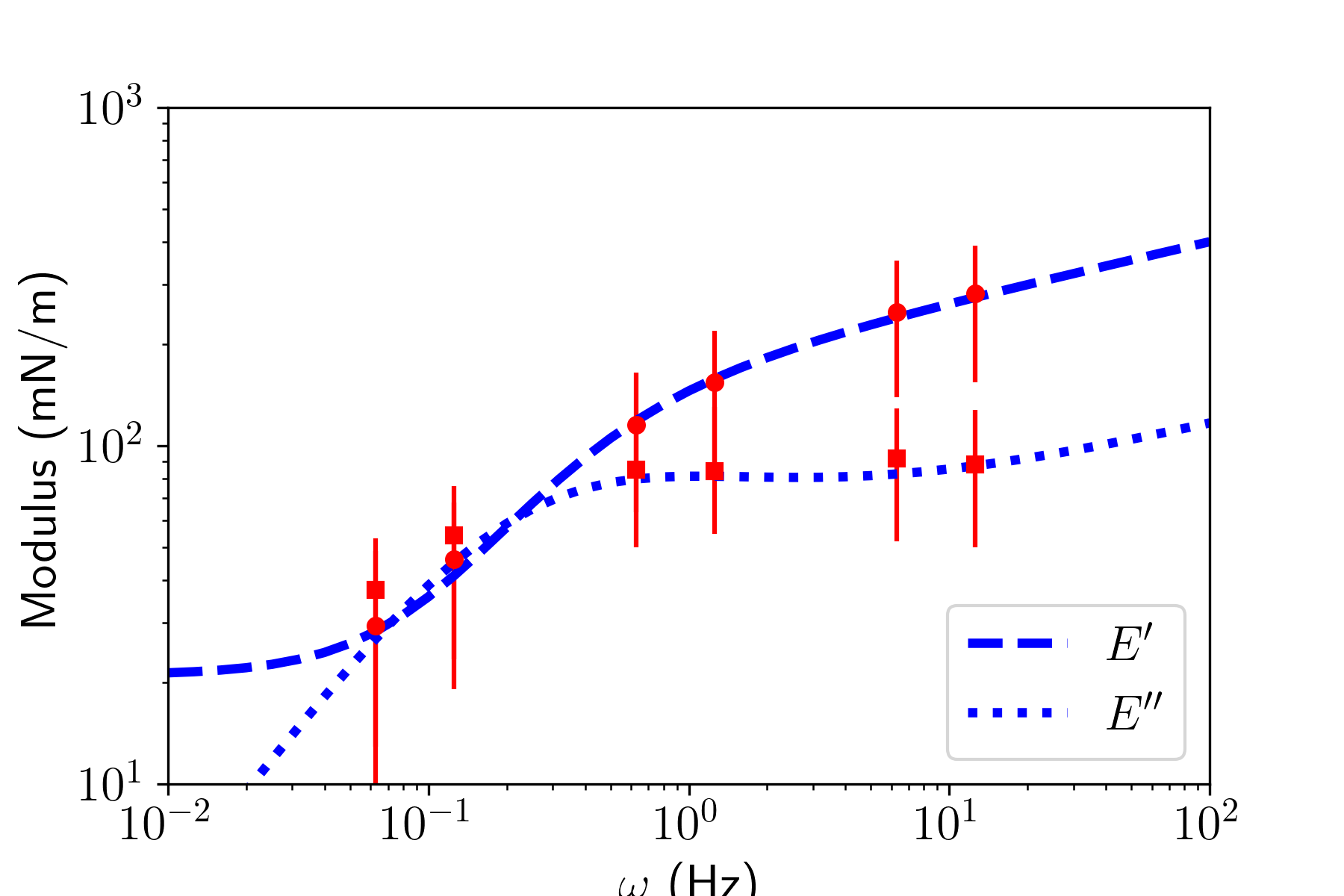}

\caption{
  \textbf{Experimental data: cortex rheology}.
  \textbf{a.} Experimental data points from \cite{FF2016} (mN/m),
  represented by the red symbols and their error bars, 
  are consistent with Eq.~\eqref{eq:def:e:withoutk} (contractile
  visco-fractional cortex rheology), with the parameter values:
$\cbetadd =  60 \, \mathrm{mN \cdot s^B /m}$,
$\etadd =  250 \, \mathrm{mN \cdot s/m}$,
$\betadd  = 0.19 $ 
(black lines).
We observe a low-frequency viscous limit, and a high frequency fractional
limit, with exponent $B$, dominated by the conservation modulus.
\textbf{b.} Red symbols and their error bars are obtained from data
plotted in panel \textbf{a} using the mapping
(\ref{eq:K:s0:g:k:L:H}-\ref{eq:nu:s0:g:k:L:H})
with  $\Psi = 1$, $\soh = \sol =  0.3 \, \mathrm{mN \, m^{-1}}$,
$\nudd = 0.41$
and $g = (1-\nudd)\,k/(1+\nudd)$
(blue lines).
They are consistent with Eq.~\eqref{eq:E:def:TVEF} (tissue viscoelastic
fractional rheology), with the parameter values:
$c_{\beta} =  500 \, \mathrm{mN \cdot s^B /m}$,
$\eta =  170 \, \mathrm{mN \cdot s/m}$,
$\beta  = 0.19 $,
$K_0 =  21 \, \mathrm{mN/m}$. 
We observe a low-frequency elastic limit, and a high frequency fractional
limit, with exponent $\beta$, dominated by the conservation modulus.
Dashed line: conservation modulus; dotted line: loss modulus. 
}
\label{fig:2D:cortexdata}
\end{figure}

In the high frequency limit 
\red{$\omega \to \infty$, 
if the cortex moduli}
$\gl$, $\kl$, $\gh$ and $\kh$ 
\red{behave elastically,
then the monolayer moduli are provided by the full expressions
of Eqs.~(\ref{eq:K:s0:g:k:L:H}-\ref{eq:nu:s0:g:k:L:H}).
} 
\red{However, if they are non-elastic in this limit,
hence if their amplitudes go to infinity,}
we obtain:
\begin{eqnarray}
\Ktis_\infty &\red{\sim}& 2\kh +\Psi\,( \kl +9\gl) \,,
\label{eq:Ktis:high:freq}
  \\
  \Gtis_\infty &\red{\sim}& 2\gh \,,
\label{eq:Gtis:high:freq}                   
\\
\Etis_\infty &\red{\sim}& \frac{4 \left( 2\gh \right) \left(  2\kh 
+\Psi\,( \kl +9\gl)\right)}{ 2\gh + 2\kh 
                 +\Psi\,( \kl +9\gl) } \,,
\label{eq:Etis:high:freq}
\\
\nutis_\infty &\red{\sim}& \frac{2\kh +\Psi\,( \kl +9\gl) -2\gh}
                  {2\kh +\Psi\,( \kl +9\gl) +2\gh} \,.
\label{eq:nutis:high:freq}
\end{eqnarray}
As a consequence, the high frequency scaling behaviour of
tissue moduli is identical to that of cortex moduli:
a viscous cortex will lead to a viscous tissue;
a fractional cortex rheology will lead to a
fractional tissue rheology with the same exponent
(see Fig.~\ref{fig:2D:cortexdata} for an example). 
\red{If the cell state is altered in such a way that 
the cortical fractional exponent is changed,
we predict that the same change will display at
the tissue scale.}

Whereas the low frequency limit of the Poisson ratio  is $\nutis_0 = 1/2$,
its high frequency limit $\nutis_\infty$ 
is expected from \eqref{eq:nutis:high:freq}
to depend on specifics of the cortex moduli.

\subsection{Experimental data}
\label{sec:FF}

Cortex viscoelastic properties have been measured recently
as a function of frequency~\cite{FF2016,Mokbel2020}
in HeLa cells squeezed between two parallel plates.
The more recent work~\cite{Mokbel2020}
takes precisely into account the spatial variations
of the cortex deformation modes during cell compression, 
with some regions stretched isotropically
and other regions subjected to some degree of shear deformation.
Assuming that the rheological properties are uniform in the entire cell,
it then yields values of the cortex Poisson ratio,
found to decrease from $0.66$ to $0.17$ 
as frequency increases from $0.1$ to $10$ Hz,
with a mean value $\nudd \approx 0.41$.
This corresponds to the compression modulus, $k$, being $1.5$ to $4$ times larger than the shear modulus, $g$.
This work~\cite{Mokbel2020}, however, does not provide the values and variations
of any modulus (shear, compression, etc)
\red{separately}.
In order to map cortex to tissue rheology,
we therefore use the cortex complex modulus data points
from the older work 
\red{in the same team}%
~\cite{FF2016} 
(see Fig~\ref{fig:2D:cortexdata}a),
although the values thus obtained 
are a mixture of shear and compression moduli.
More precisely, the compression force is interpreted
as if the tension were isotropic and uniform throughout the cortex during compression.
As a consequence, the resulting value for the modulus
would exactly match the compression modulus
in the case of a vanishing shear modulus.
For simplicity, we assume it represents the compression modulus
when comparing with our results in Section~\ref{sec:K:G:from:k:g:so:H:L}.
We have checked that, by contrast, if it is assumed it represents the shear modulus,
our results presented below are unchanged qualitatively, although fitted parameter values are altered.

In Fig~\ref{fig:2D:cortexdata}, we choose a frequency-independent cortex Poisson ratio
equal to the average value measured by~\cite{Mokbel2020}, $\nudd = 0.41$.
We also assume that the cells are cuboidal, $\Psi = 1$.
Using a value of all cortex tensions 
$\soh = \sol = \so = 0.3 \, \mathrm{mN \, m^{-1}}$~\cite{Salbreux2012},
and assuming that the moduli of lateral and horizontal facets are identical
($\kh=\kl=k$ and $\gh=\gl=g$),
the mapping (\ref{eq:K:s0:g:k:L:H}-\ref{eq:nu:s0:g:k:L:H})
yields values of the tissue Young modulus plotted in
Fig~\ref{fig:2D:cortexdata}b.
As advocated by Bonfanti \emph{et al.}~\cite{Bonfanti2020}, 
we fitted this data by the following form of the tissue Young's modulus:
\begin{eqnarray}
  E^{\mathrm{TVEF}}(\omega) = K_0 +
  \frac{1}{\frac{1}{i\omega \eta}+\frac{1}{c_{\beta}(i\omega)^{\beta}}} \,,
\label{eq:E:def:TVEF}
\end{eqnarray}
obtaining the parameter values
$\eta = (5 \pm 3) \, 10^2 \, \mathrm{mN \, s\, m^{-1}}$,
$c_{\beta} = (1.7 \pm 0.7) \, 10^2 \, \mathrm{mN\, m^{-1}}$,
$\beta = 0.185 \pm 0.010$,
$K_0 =  (2.1 \pm 0.8) \, 10^1\, \mathrm{mN/m}$. 
Since HeLa cells are softer and more fluid than MDCK cells, 
the values of $\eta$, $c_\beta$ and $K_0$
are at least one order of magnitude lower than obtained for \red{an} MDCK monolayer in \cite{Bonfanti2020}.
Interestingly, the tissue Poisson ratio resulting from our mapping
is approximately frequency independent, $\nutis = 0.77 \pm 0.01$.

We next fitted the cortex rheological data plotted in
Fig~\ref{fig:2D:cortexdata}a with  a model  
where a dashpot (viscosity $\etadd$) is combined in series 
with a fractional element (parameters $\cbetadd$, $\betadd$):
\begin{equation}
k_{\mathrm{cortex}}(\omega) =  \frac{1}{\frac{1}{i\omega \etadd} +
\frac{1}{\cbetadd (i\omega)^{\betadd}}} \,,
\label{eq:def:e:withoutk}
\end{equation}
obtaining the parameter value estimates
$\etadd = (2.5 \pm 1.6) \, 10^2 \, \mathrm{mN \, s\, m^{-1}}$,
$\cbetadd = (6.0 \pm 2.5) \, 10^1 \, \mathrm{mN\, m^{-1}}$,
$\betadd = 0.19 \pm 0.01$. 
Note that Bonfanti \textit{et al.}~\cite{Bonfanti2020} performed fits 
of the same data assuming that this rheological diagram was in addition 
in parallel with a spring of stiffness $\kdd$,
at variance with the fluid behaviour expected at long times 
for the cell cortex.
Interestingly, our results allow to map a cortex rheology
consistent with Eq.~\eqref{eq:def:e:withoutk} into a tissue rheology
consistent with Eq.~\eqref{eq:E:def:TVEF}, and to explain the change
of magnitude of the corresponding parameter values, with a tissue both
stiffer 
($c_{\beta} \simeq 3 \, \cbetadd$), and more viscous ($\eta \simeq 2 \, \etadd$),
than the cell cortex.

\subsection{Examples}

\subsubsection{Fractional cortex rheology}
\label{sec:fractional:cortex:rheology}

\red{
The monolayer moduli shown in Fig~\ref{fig:2D:cortexdata}b,
are mapped (through Eqs.~\ref{eq:K:s0:g:k:L:H}-\ref{eq:nu:s0:g:k:L:H})
from a cortex described by Eq.~\eqref{eq:def:e:withoutk}
combined with a rest tension $\so$.
They appear to display an intermediate frequency regime.
As in the tissue-scale expression~\eqref{eq:E:def:TVEF}
by Bonfanti \emph{et al.}~\cite{Bonfanti2020}, 
this regime reflects the viscosity component in the cortex rheology.
It can be shown that it is no more discernable
whenever the viscosity is high enough, namely:
} 
\begin{equation}
\red{
\etadd \gg \cbetadd^{1/\betadd}
\,\so^{1-1/\betadd}\,.
} 
\label{eq:eta:condition:cvfc}
\end{equation}
\red{
In the limit of Eq.~\eqref{eq:eta:condition:cvfc},
the cortex model of Eq.~\eqref{eq:def:e:withoutk}
can be simplified into a simple fractional rheology,
which we now explore.
} 

\red{Let} 
all facet moduli 
($\kh$, $\kl$, $\gh$, $\gl$) 
have the same dependence on frequency,
for instance fractional elements with exponent $\beta$:
\begin{eqnarray}
\kh(\omega) &=& \kh_\beta\,(i\,\omega)^\beta \,,
\\
\kl(\omega) &=& \kl_\beta\,(i\,\omega)^\beta \,,
\\
\gh(\omega) &=& \gh_\beta\,(i\,\omega)^\beta \,,
\\
\gl(\omega) &=& \gl_\beta\,(i\,\omega)^\beta \,,
\end{eqnarray}
where the prefactors $\kh_\beta$, etc, are real numbers.
\red{These expressions imply
that the cortex Poisson ratios
for horizontal and lateral facets,
given by Eq.~\eqref{eq:nudd:k:g},
are frequency-independent.
} 

The monolayer moduli $\Ktis$ and $\Gtis$,
given by Eqs.~(\ref{eq:K:s0:g:k:L:H},%
\ref{eq:G:s0:g:k:L:H}),
can then be approximated at both low and high frequencies
by the respective sums of their asymptotic expressions in both limits:
\begin{eqnarray}
  \Ktis
  &\simeq& 3\soh + \Ktis_\beta\,(i\,\omega)^\beta \,,
\label{eq:Ksimpl}
\\
  \Gtis
  &\simeq& \soh + 2\gh_\beta\,(i\,\omega)^\beta \,,
\label{eq:Gsimpl}
\end{eqnarray}
where $\Ktis_\beta = 2\kh_\beta +\Psi\,( \kl_\beta +9\gl_\beta)$.

If all cortex moduli are fractional,
$K$ and $G$ both evolve from elastic at low frequencies
(depending only on the cortex tension at rest, $\soh$)
to fractional at large frequencies.
The crossovers at intermediate frequencies
depend on the details of the parameter values.
This will in particular be the case 
for a purely viscous cortex ($\beta=1$).

\begin{figure}[ht!]
\centering
\textbf{a.}
\includegraphics[scale=0.4]{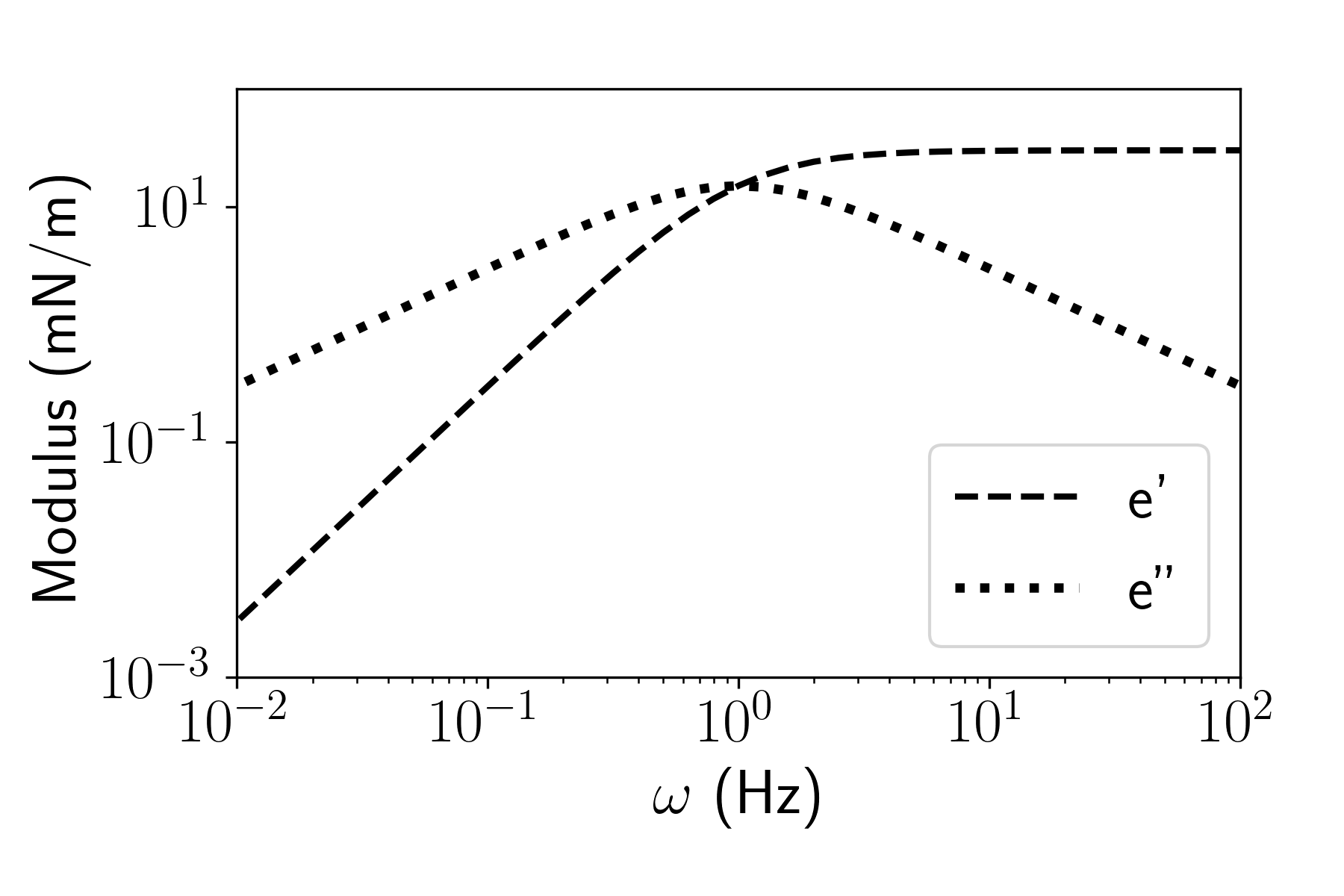}
\textbf{b.}
\includegraphics[scale=0.4]{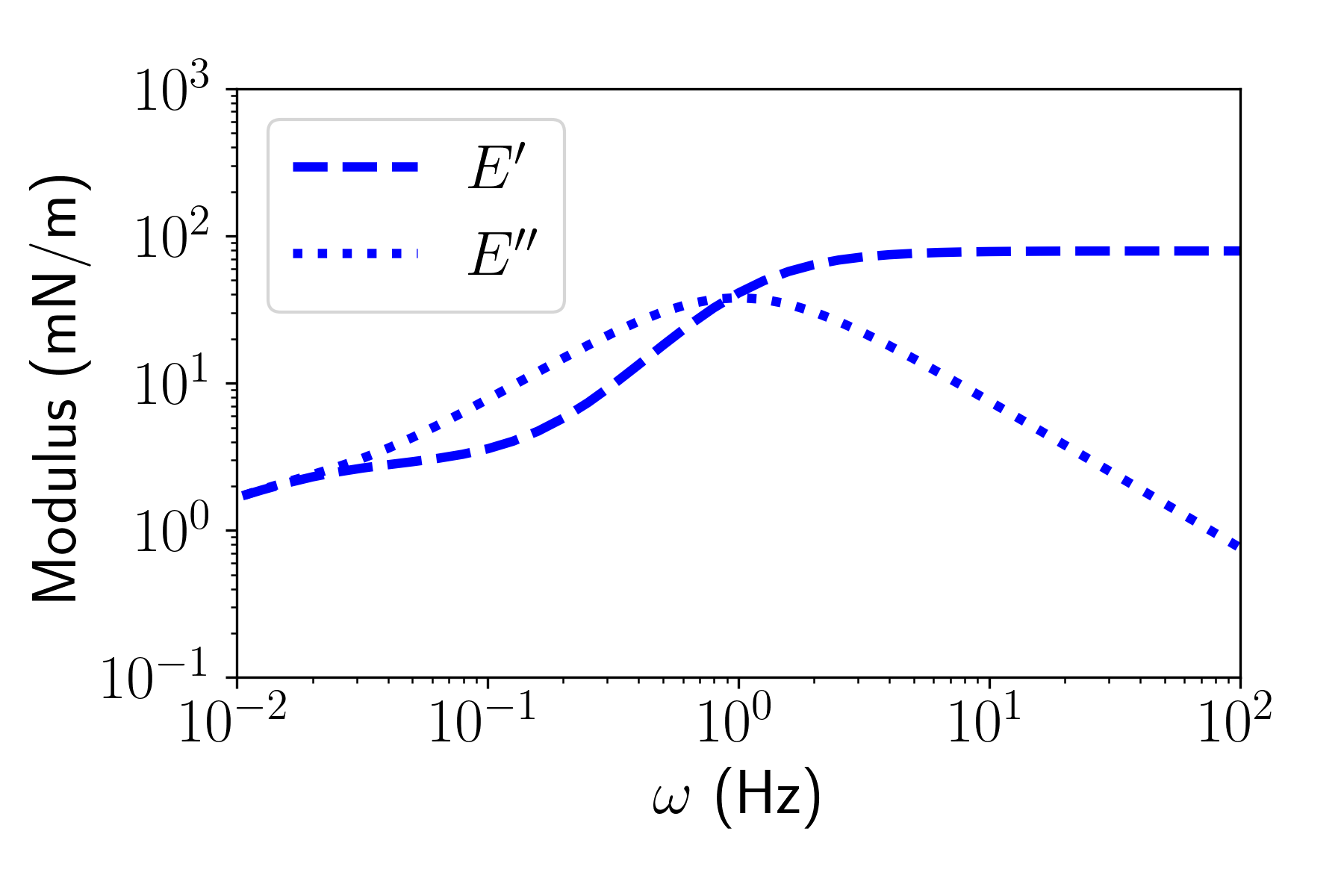}
\caption{
  \textbf{Tissue moduli based on a Maxwell cortex rheology}.
  Dashed lines: conservation moduli; dotted lines: loss moduli (mN/m).
\textbf{a.} Cortex rheology, Eq.~\eqref{eq:e:CVEC}, with parameters
$\eta = 30 \, \mathrm{mN \cdot s/m}$ and
$g = 30 \, \mathrm{mN /m}$.
\textbf{b.} Tissue rheology, obtained using
Eqs.~(\ref{eq:K:s0:g:k:L:H},%
\ref{eq:G:s0:g:k:L:H}) with parameters $\nu = 0.4$, $\Psi = 1$ and 
 $\sigma_0  = 0.3 \, \mathrm{mN /m}$.
 We obtain a low-frequency (not shown) and a high-frequency 
elastic limit, with an intermediate dissipative regime.
}
\label{fig:hexagones:tissu_avec_cortexCVEC}
\end{figure}

\subsubsection{Maxwell cortex rheology}

\red{
Another (this time, elastic) limit of the  model \eqref{eq:def:e:withoutk},
with $\beta\rightarrow 0$,
leads to a
}
Maxwell model for the cortex,
with a viscous/elastic behaviour respectively in the
low/high frequency limit.
Such a model has been proposed for the cell cortex rheology
in~\cite{Etienne2015}, and reads
\begin{eqnarray}
e^{\mathrm{Maxwell}} =  \frac{1}{\frac{1}{i\omega \eta}+\frac{1}{g}} \,,
\label{eq:e:CVEC}
\end{eqnarray}
(see Fig.~\ref{fig:hexagones:tissu_avec_cortexCVEC}a).
In this case, the high-frequency tissue rheological behaviour is elastic
(see Fig.~\ref{fig:hexagones:tissu_avec_cortexCVEC}b),
at variance with the power-law behaviour observed experimentally
in this limit \cite{Bonfanti2020}.
This model, introduced by~\cite{Etienne2015},
constitutes an interesting heuristic proposal
for the cortex mechanics,
which incorporates in the most simple manner
the ingredients of cortex active tension ($\so$),
elasticity ($g$) due to the actin fiber network,
and relaxation (here with a unique timescale $\eta/g$)
due to monomer and crosslinker stochastic renewal.
It has been applied successfully 
to the microplate single cell 
experiment~\cite{Desprat2005,Balland2006}.
A similar model
has been used to account 
for experimental observations
of the effects of single cortex
manipulations~\cite{Bambardekar2014}.

\section{Inverse mapping of tissue (macro-)rheology
  to cortex (micro-)rheology}
\label{sec:inverse:mapping}

\subsection{Low frequency inversion}

The low frequency moduli of the monolayer
are real numbers according to
expressions~(\ref{eq:Ktis:low:freq}-\ref{eq:Etis:low:freq}).
In other words, the monolayer is purely elastic in this limit,
with a stiffness proportional to the rest tension
of the horizontal cortices, $\soh$.
This feature allows for a direct access
to this cell-scale quantity through macroscopic measurements.
The rest tension of the lateral facets, $\sol$,
is then derived from the cell aspect ratio, $\Psi$,
as a result of Eq.~\eqref{eq:def:Psi}:
\begin{eqnarray}
\soh &=& \Gtis_0 = \frac{\Ktis_0}{3} = \frac{\Etis_0}{3} \,,
\label{eq:inversion:low:freq:soh}
\\
\sol &=& \frac{\soh}{\Psi} \,.
\label{eq:inversion:low:freq:sol}
\end{eqnarray}
However, the cortical tension of lateral sides $\sol$, 
being absent from both the low and high frequency limits
cannot be deduced from such measurements.

In practice, tissues flow over timescales large compared to the typical rearrangement/cell division/cell extrusion times.
In \cite{Tlili2020}, the viscoelastic time of a MDCK monolayer 
on a flat substrate was measured to be of the order of 1 hour.
The measurement of the "low frequency" moduli $\Gtis_0$,
$\Ktis_0$ and $\Etis_0$ would need to be performed 
over time scales shorter than this viscoelastic time.

\subsection{No general inversion}
\label{sec:no:general:inversion}

The most general goal,
when attempting to inverse the mapping obtained in Section~\ref{sec:K:G:from:k:g:so:H:L},
would be the following.
Let us assume that sophisticated mechanical measurements
provide two independent complex moduli
as a function of the angular frequency,
for instance: $\Ktis(\omega)$ and $\Gtis(\omega)$,
from which we subtract the low frequency limits:
\begin{eqnarray}
\DKtis(\omega) &=& \Ktis(\omega) -\Ktis_0 \,,
\label{eq:def:DKtis}
\\
\DGtis(\omega) &=& \Gtis(\omega) -\Gtis_0 \,.
\label{eq:def:DGtis}
\end{eqnarray}
Once the rest tensions
are known from the low frequency analysis,
see Eqs.~(\ref{eq:inversion:low:freq:soh},%
\ref{eq:inversion:low:freq:sol}),
the goal is to obtain
two independent complex moduli
of both the horizontal and the lateral facets:
$\kh(\omega)$, $\gh(\omega)$, $\kl(\omega)$ and $\gl(\omega)$.

It is clearly not possible,
with no additional assumption,
to uniquely determine four such quantities
from only both monolayer scale quantities
$\DKtis(\omega)$ and $\DGtis(\omega)$.
Below, we review some assumption examples
and the corresponding expressions of the cortex scale quantities.

\subsection{Inversion under tension scaling assumption}
\label{sec:inversion:tension:scaling}

In order to go beyond this impossibility,
let us now assume that the phenomenon
that causes the tension of the horizontal and lateral facets to differ
equally applies to the complex moduli.
In other words, let us assume that the tension ratio, $\Psi$,
also applies to the moduli, at all frequencies:
\begin{eqnarray}
\frac{\soh}{\sol} = \Psi
&=& \frac{\kh(\omega)}{\kl(\omega)}
= \frac{\gh(\omega)}{\gl(\omega)} \,.
\label{eq:hyp:tension:scaling}
\end{eqnarray}
If $\Psi = 1$ in addition to this scaling assumption,
the rheological properties of all facets would become identical.

With assumption \eqref{eq:hyp:tension:scaling},
the expressions of $\Ktis$ and $\Gtis$,
given by Eqs.~(\ref{eq:K:s0:g:k:L:H},%
\ref{eq:G:s0:g:k:L:H})
can be rewritten in terms of $\DKtis$ and $\DGtis$,
see Eqs.~(\ref{eq:def:DKtis},\ref{eq:def:DGtis}).
Using the assumption expressed above
as Eq.~\eqref{eq:hyp:tension:scaling},
they can be rewritten in terms of only 
$\soh$, $\kh$ and $\gh$, 
with no influence of $\Psi$:
\begin{eqnarray}
\DKtis &=& 3\kh +9\gh \,,
\label{eq:DK:s0:gh:kh}
\\
\DGtis &=& 2\gh
+\frac{1}
{\frac{1}{2 \soh} +\frac{1}{2\kh +2\gh}} \,.
\label{eq:DG:s0:gh:kh}
\end{eqnarray}

Assuming that not only $\soh$,
but also $\DKtis$ and $\DGtis$ are positive real numbers,
one can invert these equations
and obtain the values of the cortex moduli
of the horizontal facets:
\begin{eqnarray}
\kh &=& \frac{1}{4}\left(
3\soh  -\sqrt{D} +\DKtis -3\DGtis
\right) \,,
\label{eq:kh:soh:D:DK:DG}
\\
\gh &=& \frac{1}{12}\left(
\sqrt{D} -3\soh +\DKtis +3\DGtis
\right) \,,
\label{eq:gh:soh:D:DK:DG}
\\
D &=& 9\soh^2 +18\soh\,(\DKtis-3\DGtis) +(\DKtis -3\DGtis)^2 \,.
\end{eqnarray}
The moduli of the lateral facets 
are then immediately obtained using the cell aspect ratio:
\begin{eqnarray}
\kl &=& \kh / \Psi \,,
\\
\gl &=& \gh / \Psi \,.
\end{eqnarray}
In Eqs.~(\ref{eq:kh:soh:D:DK:DG},\ref{eq:gh:soh:D:DK:DG}),
the sign in front of $\sqrt{D}$
was chosen so as to satisfy the low frequency limit.
The quantities $\DKtis(\omega)$ and $\DGtis(\omega)$
are in fact complex numbers,
and one may assume that, as combinations
of springs ($\beta=0$), dashpots ($\beta=1$)
and fractional elements ($0<\beta<1$),
their real and imaginary parts are both positive.
%
We shall adopt a definition of $\sqrt{D}$
that remains continuous in all accessible regions
around the low frequency (real) value $\sqrt{D}=3\soh$.

\red{Let us consider a} 
first particular situation
where the monolayer Poisson ratio $\nutis$
is equal to $1/2$ not only in the low frequency limit
but also at all frequencies.
\red{
Eq.~\eqref{eq:nu:s0:g:k:L:H}
then implies that
} 
$\Ktis(\omega) = 3\Gtis(\omega)$.
\red{Hence, we have} 
$\DKtis(\omega)=3\DGtis(\omega)$.
\red{As a consequence,} 
$\sqrt{D}=3\soh$ is constant.
The above results then suggest
that, ignoring the influence of their rest tension $\soh$,
the cortices' bulk moduli vanish, 
while their shear moduli are directly related to the monolayer shear modulus:
\begin{eqnarray}
\Psi\kl = \kh &=& 0 \,,
\\
\Psi\gl = \gh &=& \frac{1}{2}\,\DGtis
= \frac{1}{2}\,(\Gtis -\Gtis_0) \,.
\end{eqnarray}

In a second particular situation, we
assume not only that horizontal and lateral facet moduli
are in the same ratio as the rest tension, 
see Eq.~\eqref{eq:hyp:tension:scaling},
which implies that all facets 
have identical Poisson's ratio $\nudd(\omega)$,
but we also assume that this value is constant:
$\nudd(\omega)=\mathrm{const}$.
It is then more convenient
to express the cortex compression and shear moduli, $k$ and $g$,
in terms of the cortex Young modulus $\Edd$ and Poisson ratio $\nudd$:
\begin{eqnarray}
\kh &=& \frac{\Eddh}{2\,(1-\nudd)} \,,
\label{eq:kh:e:nuc}
\\
\gh &=& \frac{\Eddh}{2\,(1+\nudd)} \,.
\label{eq:gh:e:nuc}
\end{eqnarray}
Eqs.~(\ref{eq:K:s0:g:k:L:H},\ref{eq:G:s0:g:k:L:H}) then become:
\begin{eqnarray}
\Ktis &=& 
3\,\soh 
+\frac{6-3\nudd}{1-\nudd^2}\,\Eddh \,,
\label{eq:K:s0:e:nuconst}
\\
\Gtis &=& 
\soh+\frac{1}{1+\nudd}\,\Eddh
+\frac{1}
{
\frac{1}{2 \soh} +\frac{1-\nudd^2}{2\,\Eddh}
} \,,
\label{eq:G:s0:e:nuconst}
\end{eqnarray}
while the monolayer Young modulus $\Etis$ 
and Poisson ratio $\nutis$ 
are then obtained
from Eqs.~(\ref{eq:E:s0:g:k:L:H}, \ref{eq:nu:s0:g:k:L:H}).
That implies that the cortex moduli
can be immediately expressed from the monolayer bulk modulus $\Ktis$:
\begin{eqnarray}
\Eddh &=& 
\frac{1-\nudd^2}{6-3\nudd}\,\DKtis \,,
\label{eq:e:DK:s0:nuconst}
\\
\kh &=& 
\frac{1+\nudd}{12-6\nudd}\,\DKtis \,,
\label{eq:kh:DK:s0:nuconst}
\\
\gh &=& 
\frac{1-\nudd}{12-6\nudd}\,\DKtis \,.
\label{eq:gh:DK:s0:nuconst}
\end{eqnarray}
This particular case illustrates how,
when $\nudd(\omega)=\mathrm{const}$,
cortex and tissue rheology 
may be identical, as suggested both by \cite{Bonfanti2020} and 
Fig.~\ref{fig:2D:cortexdata}.

An expression for the monolayer shear modulus $\Gtis$ can also be obtained, 
for instance for the cortex shear modulus $\gh$:
\begin{eqnarray}
4\gh &=& 
\DGtis-(3-\nudd)\soh 
+\sqrt{(\DGtis-(3-\nudd)\soh)^2
+(1-\nudd)\soh\DGtis} \,.
\label{eq:e:DG:s0:nuconst}
\end{eqnarray}
Eqs.~(\ref{eq:kh:e:nuc}-\ref{eq:gh:e:nuc})
then readily yield expressions for the Young modulus $\Eddh$ 
and compression modulus $\kh$.
Eq.~\eqref{eq:e:DG:s0:nuconst}
shows that a strong similarity is expected between the frequency dependencies
of the (micro-scale) cortex moduli 
and the (macro-scale) monolayer shear modulus.
If in addition, $\nudd = 1$
(in other words, a possibly unrealistic assumption of a 2d-incompressible cortex), 
then the tissue becomes incompressible ($\Ktis \to \infty$), 
Eq.~\eqref{eq:e:DG:s0:nuconst} becomes
\begin{eqnarray}
2\gh &=& 
\DGtis-2\soh  \,.
\label{eq:e:DG:s0:nu1}
\end{eqnarray}
If we rather assume $\nudd = 0$
as in~\cite{Etienne2015}, 
Eqs.~(\ref{eq:e:DK:s0:nuconst}-\ref{eq:gh:DK:s0:nuconst}) become
$\Eddh=2\gh=2\kh =\DKtis/6$.

\section{Discussion and perspectives}
\label{sec:discussion:perspectives}

To summarize, we characterize in this work
the mechanical behaviour
of an epithelial tissue by establishing the link between 
the mechanics at the cytoskeletal scale and the  mechanics 
at the epithelial scale. We describe theoretically the 
linear  rheology  of an ordered assembly of hexagonal cells, 
as a function of those of the cell  cortices. We also discuss,
in particular cases, the inverse problem that starts from 
the epithelial rheology and deduces the cortical rheology. 
\red{In the low-frequency limit, we obtain that the monolayer is elastic,
with moduli proportional to the cortex rest tensions, as expected
of tensegrity structures.
In other frequency ranges,
the monolayer rheology reflects the main features
of the cortex rheologies.}

We hope that this work  will be conducive to a better
understanding of the contribution of 
cellular constituents in the epithelial mechanical behaviour.
It suggests that the 
rich rheological behaviour of cell cortices
should be taken into account when formulating
(disordered) vertex models, beyond the current,
standard definition of an energy function based on 
constant  cortex (cell-cell junction) tensions.

In order to test whether the full 3D geometry 
depicted on Fig.~\ref{fig:ingredients}a
is required to predict the monolayer rheology,
we attempted to reduce the dimensionality
and considered a regular tiling 
by \textit{planar} hexagonal cells. As shown in Appendix~\ref{sec:2d},
this geometry leads to an elastic high-frequency rheology,
at variance with experimental observations.
The results of these calculations suggest that, at least for ordered tilings, the 
rheological behaviour of lateral cell cortices 
and their mechanical coupling with horizontal cell cortices
cannot be ignored when determining the tissue rheology.

Our results are based on several simplifying
assumptions that may be relaxed in the future, for instance considering 
the effect of disorder (whose impact may be assessed by numerical simulations)~\cite{Kruyt2007};
an asymmetry in apico-basal tension or moduli~\cite{Messal2019,Fouchard2020}; 
a realistic bulk cell rheologyl~\cite{Verdier2009}
or different boundary conditions~\cite{Harris2013}. 
In the experiments~\cite{Wyatt2019}
a finite initial tension of the monolayer is observed in the $x$-direction.
Moreover, an actin cable is present
along each free, lateral edge of the suspended monolayer.
These edges being slightly curved, they probably exert some tension on the monolayer in the $y$-direction.
The initial monolayer tension seems isotropic in the $xy$-plane
(see Fig.~2a in~\cite{Wyatt2019}).
We have here neglected any such initial monolayer tension.
While we focused on the planar geometry characteristic of
cell monolayers, a natural extension would be 
to represent a 3D tissue (for instance a cellular spheroid)
by a tiling composed of Kelvin cells,
for which the low frequency, elastic limit
has been established in the field of liquid foams~\cite{Reynelt1993}.

In this work, the fact that horizontal facets are assumed flat
is by itself a simplification.
Indeed, force balance implies that a right angle at the edge 
between lateral and horizontal facets
would correspond to horizontal tension values 
much larger than lateral tension values.
A more realistic geometry,  with curved apical and basal facets,
is left for future studies.

Perhaps more importantly, the cells have more than two 
in-plane degrees of freedom
and they need to obey force balance on vertical edges, 
see Eq.~\eqref{eq:force:balance:six:cortex:tensions}.
Hence, even though the rectangular representative volume
has to deform exactly in the same proportions as the monolayer,
the cells in general do not deform homogeneously,
instead they contain some elements
that move in a non-affine manner \red{(see Appendix~\ref{sec:affinity})}.
Here, however, in order to simplify the description,
we calculate the stress in the horizontal facets
as if their deformation were homogeneous
(and thus identical to the monolayer in-plane deformation).

Implementing and validating the inverse mapping from monolayer
to cell cortex rheology, depends practically on the availability of 
the corresponding experimental measurements.
A first valuable contribution
would be to test the monolayer
in such a way as to measure two independent macroscopic moduli.
A second milestone would be 
to have access to measurements of the cortex rheology of cells within a monolayer
rather than of isolated single cells,
as pioneered by \cite{Clement2017,Esfahani2021,Pietuch2013}.
That may then help assessing the scaling assumption Eq.~\eqref{eq:hyp:tension:scaling}
put forward in Section~\ref{sec:inversion:tension:scaling}.

We hope that this work will foster more experimental work, 
measuring both cortex and tissue rheology, 
in view of the direct and inverse mappings that link them.

\section{Acknowledgements}

We acknowledge fruitful discussions with Anne Tanguy, \red{Magali Suzanne and Camille No\^us, as well as} 
with Jocelyn \'Etienne whose cortex rheology model\red{~\cite{Etienne2015}}
has been a central source of inspiration for the present work.
We thank Jonathan Fouchard and François Graner for a careful 
reading of the manuscript.
\red{
We also thank the referees for their comments,
which helped clarify several aspects.
} 

\bibliographystyle{unsrt} 
\bibliography{etienne}

\begin{thebibliography}{10}

\bibitem{Heisenberg2013}
Carl-Philipp Heisenberg and Yohanns Bellaïche.
\newblock Forces in tissue morphogenesis and patterning.
\newblock {\em Cell}, 153:948--962, 2013.

\bibitem{Goodwin2021}
Katharine Goodwin and Celeste~M. Nelson.
\newblock Mechanics of development.
\newblock {\em Developmental Cell}, 56:240--250, 2022.

\bibitem{Valet2021}
Manon Valet, Eric~D. Siggia, and Ali~H. Brivanlou.
\newblock Mechanical regulation of early vertebrate embryogenesis.
\newblock {\em Nature Reviews Molecular Cell Biology}, 23:169--184, 2021.

\bibitem{Verdier2009}
Claude Verdier, Jocelyn Etienne, Alain Duperray, and Luigi Preziosi.
\newblock Rheological properties of biological materials.
\newblock {\em C.R. Physique}, 10:790–811, 2009.

\bibitem{Petridou2019}
Nicoletta~I Petridou and Carl-Philipp Heisenberg.
\newblock Tissue rheology in embryonic organization.
\newblock {\em The {EMBO} Journal}, 38:e102497, 2019.

\bibitem{Guevorkian2010prl}
Karine Guevorkian, Marie-Jos\'ee Colbert, M\'elanie Durth, Sylvie Dufour, and
  Fran\c{c}oise Brochard-Wyart.
\newblock Aspiration of biological viscoelastic drops.
\newblock {\em Phys. Rev. Lett.}, 104:218101, 2010.

\bibitem{stirbat2013}
Tomita~Vasilica Stirbat, Sham Tlili, Thibault Houver, Jean-Paul Rieu, Catherine
  Barentin, and Hélène Delanoë-Ayari.
\newblock Multicellular aggregates: a model system for tissue rheology.
\newblock {\em Eur Phys J E Soft Matter}, 36:9898, 2013.

\bibitem{Mary2022}
Ga\"etan Mary, Fran\c{c}ois Mazuel, Vincent Nier, Florian Fage, Ir\`ene Nagle,
  Louisiane Devaud, Jean-Claude Bacri, Sophie Asnacios, Atef Asnacios, Cyprien
  Gay, Philippe Marcq, Claire Wilhelm, and Myriam Reffay.
\newblock All-in-one rheometry and nonlinear rheology of multicellular
  aggregates.
\newblock {\em Phys. Rev. E}, 2022.

\bibitem{Vincent2015}
Romaric Vincent, Elsa Bazellières, Carlos Pérez-González, Marina Uroz,
  Xavier Serra-Picamal, and Xavier Trepat.
\newblock Active tensile modulus of an epithelial monolayer.
\newblock {\em Phys Rev Lett}, 115:248103, 2015.

\bibitem{Nier2018}
Vincent Nier, Gr{\'{e}}goire Peyret, Joseph d'Alessandro, Shuji Ishihara,
  Benoit Ladoux, and Philippe Marcq.
\newblock Kalman inversion stress microscopy.
\newblock {\em Biophysical Journal}, 115(9):1808--1816, 2018.

\bibitem{Tlili2020}
S.~Tlili, M.~Durande, C.~Gay, B.~Ladoux, F.~Graner, and H.~Delano\"e-Ayari.
\newblock Migrating epithelial monolayer flows like a maxwell viscoelastic
  liquid.
\newblock {\em Phys. Rev. Lett.}, 125:088102, 2020.

\bibitem{Harris2013}
Andrew~R Harris, Julien Bellis, Nargess Khalilgharibi, Tom Wyatt, Buzz Baum,
  Alexandre~J Kabla, and Guillaume~T Charras.
\newblock Generating suspended cell monolayers for mechanobiological studies.
\newblock {\em Nat Protoc}, 8:2516--2530, 2013.

\bibitem{Khalilgharibi2019}
Nargess Khalilgharibi, Jonathan Fouchard, Nina Asadipour, Ricardo Barrientos,
  Maria Duda, Alessandra Bonfanti, Amina Yonis, Andrew Harris, Payman Mosaffa,
  Yasuyuki Fujita, Alexandre Kabla, Yanlan Mao, Buzz Baum, Jos{\'{e}}~J
  Mu{\~{n}}oz, Mark Miodownik, and Guillaume Charras.
\newblock Stress relaxation in epithelial monolayers is controlled by the
  actomyosin cortex.
\newblock {\em Nature Physics}, 15:839–847, 2019.

\bibitem{Bonfanti2020}
A.~Bonfanti, J.~Fouchard, N.~Khalilgharibi, G.~Charras, and A.~Kabla.
\newblock A unified rheological model for cells and cellularised materials.
\newblock {\em Royal Society Open Science}, 7:190920, 2020.

\bibitem{Pullarkat2007}
Pramod~A. Pullarkat, Pablo~A. Fernández, and Albrecht Ott.
\newblock Rheological properties of the eukaryotic cell cytoskeleton.
\newblock {\em Phys. Rep.}, 449:29--53, 2007.

\bibitem{Fabry2001}
\red{Ben Fabry}, Geoffrey~N. Maksym, James~P. Butler, Michael Glogauer, Daniel
  Navajas, and Jeffrey~J. Fredberg.
\newblock Scaling the microrheology of living cells.
\newblock {\em Phys. Rev. Lett.}, 87:148102, 2001.

\bibitem{Lenormand2004}
\red{Guillaume Lenormand}, Emil Millet, Ben Fabry, James~P. Butler, and
  Jeffrey~J. Fredberg.
\newblock Linearity and time-scale invariance of the creep function in living
  cells.
\newblock {\em Journal of The Royal Society Interface}, 1(1):91--97, nov 2004.

\bibitem{Desprat2005}
Nicolas Desprat, Alain Richert, Jacqueline Simeon, and Atef Asnacios.
\newblock Creep function of a single living cell.
\newblock {\em Biophys. J.}, 88:2224--2233, 2005.

\bibitem{Balland2006}
Martial Balland, Nicolas Desprat, Delphine Icard, Sophie Féréol, Atef
  Asnacios, Julien Browaeys, Sylvie Hénon, and François Gallet.
\newblock Power laws in microrheology experiments on living cells: Comparative
  analysis and modeling.
\newblock {\em Phys. Rev. E}, 74:021911, 2006.

\bibitem{Chugh2018}
Priyamvada Chugh and Ewa~K. Paluch.
\newblock The actin cortex at a glance.
\newblock {\em Journal of Cell Science}, 131:jcs186254, 2018.

\bibitem{Salbreux2012}
Guillaume Salbreux, Guillaume Charras, and Ewa Paluch.
\newblock Actin cortex mechanics and cellular morphogenesis.
\newblock {\em Trends in Cell Biology}, 22:536 -- 545, 2012.

\bibitem{FF2016}
Elisabeth Fischer-Friedrich, Yusuke Toyoda, Cedric~J. Cattin, Daniel~J.
  M\"uller, Anthony~A. Hyman, and Frank J\"ulicher.
\newblock Rheology of the active cell cortex in mitosis.
\newblock {\em Biophysical Journal}, 111:589 -- 600, 2016.

\bibitem{Mokbel2020}
Marcel Mokbel, Kamran Hosseini, Sebastian Aland, and Elisabeth
  Fischer-Friedrich.
\newblock The poisson ratio of the cellular actin cortex is frequency
  dependent.
\newblock {\em Biophysical Journal}, 118:1968--1976, 2020.

\bibitem{Clement2017}
Raphaël Cl{\'{e}}ment, Beno{\^{\i}}t Dehapiot, Claudio Collinet, Thomas
  Lecuit, and Pierre-Fran{\c{c}}ois Lenne.
\newblock Viscoelastic dissipation stabilizes cell shape changes during tissue
  morphogenesis.
\newblock {\em Current Biology}, 27:3132--3142.e4, 2017.

\bibitem{Esfahani2021}
Amir~Monemian Esfahani, Jordan Rosenbohm, Bahareh~Tajvidi Safa, Nickolay~V.
  Lavrik, Grayson Minnick, Quan Zhou, Fang Kong, Xiaowei Jin, Eunju Kim, Ying
  Liu, Yongfeng Lu, Jung~Yul Lim, James~K. Wahl, Ming Dao, Changjin Huang, and
  Ruiguo Yang.
\newblock Characterization of the strain-rate{\textendash}dependent mechanical
  response of single cell{\textendash}cell junctions.
\newblock {\em Proceedings of the National Academy of Sciences},
  118:e2019347118, 2021.

\bibitem{Pietuch2013}
Anna Pietuch, Bastian~Rouven Brückner, Tamir Fine, Ingo Mey, and Andreas
  Janshoff.
\newblock Elastic properties of cells in the context of confluent cell
  monolayers: impact of tension and surface area regulation.
\newblock {\em Soft Matter}, 9(48):11490, 2013.

\bibitem{Saha2016}
Arnab Saha, Masatoshi Nishikawa, Martin Behrndt, Carl-Philipp Heisenberg, Frank
  Jülicher, and Stephan W. Grill.
\newblock Determining physical properties of the cell cortex.
\newblock {\em Biophysical Journal}, 110:1421–1429, 2016.

\bibitem{Gibson2014}
Lorna~J. Gibson and Michael~F. Ashby.
\newblock {\em Cellular Solids}.
\newblock Cambridge University Press, 2014.

\bibitem{Weaire01}
D.~Weaire and S.~Hutzler.
\newblock {\em {The Physics of Foams}}.
\newblock Oxford University Press, 2001.

\bibitem{Cantat2013}
I.~Cantat, S.~Cohen-Addad, F.~Elias, F.~Graner, R.~H\"ohler, O.~Pitois,
  F.~Rouyer, and A.~Saint-Jalmes.
\newblock {\em Foams: structure and dynamics}.
\newblock Oxford University Press, ed. S.J. Cox, 2013.

\bibitem{Mead1970}
\red{J Mead}, T~Takishima, and D~Leith.
\newblock Stress distribution in lungs: a model of pulmonary elasticity.
\newblock {\em Journal of Applied Physiology}, 28(5):596--608, may 1970.

\bibitem{Stamenovic1990}
\red{D. Stamenovic}.
\newblock Micromechanical foundations of pulmonary elasticity.
\newblock {\em Physiological Reviews}, 70(4):1117--1134, oct 1990.

\bibitem{Cavalcante2005}
\red{Francisco S. A. Cavalcante}, Satoru Ito, Kelly Brewer, Hiroaki Sakai,
  Adriano~M. Alencar, Murilo~P. Almeida, Jos{\'{e}}~S. Andrade, Arnab Majumdar,
  Edward~P. Ingenito, and B{\'{e}}la Suki.
\newblock Mechanical interactions between collagen and proteoglycans:
  implications for the stability of lung tissue.
\newblock {\em Journal of Applied Physiology}, 98(2):672--679, feb 2005.

\bibitem{Fletcher2014}
Alexander~G Fletcher, Miriam Osterfield, Ruth~E Baker, and Stanislav~Y
  Shvartsman.
\newblock Vertex models of epithelial morphogenesis.
\newblock {\em Biophys J}, 106:2291--2304, 2014.

\bibitem{Alt2017}
Silvanus Alt, Poulami Ganguly, and Guillaume Salbreux.
\newblock Vertex models: from cell mechanics to tissue morphogenesis.
\newblock {\em Philosophical Transactions of the Royal Society B: Biological
  Sciences}, 372:20150520, 2017.

\bibitem{Princen1986}
H.M. Princen and A.D. Kiss.
\newblock Rheology of foams and highly concentrated emulsions: {III.} static
  shear modulus.
\newblock {\em J. Colloid Interface Sci.}, 112:427--437, 1986.

\bibitem{Reynelt1993}
D.A. Reinelt and A.M. Kraynik.
\newblock Large elastic deformations of three-dimensional foams and highly
  concentrated emulsions.
\newblock {\em J. Colloid Interface Sci.}, 159:460--470, 1993.

\bibitem{Ishihara2017}
Shuji Ishihara, Philippe Marcq, and Kaoru Sugimura.
\newblock From cells to tissue: {A} continuum model of epithelial mechanics.
\newblock {\em Physical Review E}, 96:022418, 2017.

\bibitem{Murisic2015}
Nebojsa Murisic, Vincent Hakim, Ioannis~G Kevrekidis, Stanislav~Y Shvartsman,
  and Basile Audoly.
\newblock From discrete to continuum models of three-dimensional deformations
  in epithelial sheets.
\newblock {\em Biophys J}, 109:154--163, 2015.

\bibitem{Okuda2015}
Satoru Okuda, Yasuhiro Inoue, Mototsugu Eiraku, Taiji Adachi, and Yoshiki
  Sasai.
\newblock Modeling cell apoptosis for simulating three-dimensional
  multicellular morphogenesis based on a reversible network reconnection
  framework.
\newblock {\em Biomechanics and Modeling in Mechanobiology}, 15:805--816, 2015.

\bibitem{Tong2021}
Sijie Tong, Navreeta~K. Singh, Rastko Sknepnek, and Andrej Kosmrlj.
\newblock Linear viscoelastic properties of the vertex model for epithelial
  tissues.
\newblock {\em ArXiV}, page 2102.11181, 2021.

\bibitem{Tlili2015}
Sham Tlili, Cyprien Gay, François Graner, Philippe Marcq, François Molino,
  and Pierre Saramito.
\newblock Mechanical formalisms for tissue dynamics.
\newblock {\em Eur Phys J E Soft Matter}, 38:121, 2015.

\bibitem{Grossman2021}
Doron Grossman and Jean-Fran\c{c}ois Joanny.
\newblock Rheology of 2d vertex model.
\newblock {\em ArXiV}, page 2112.04047, 2021.

\bibitem{Rauzi2008}
Matteo Rauzi, Pascale Verant, Thomas Lecuit, and Pierre-François Lenne.
\newblock Nature and anisotropy of cortical forces orienting drosophila tissue
  morphogenesis.
\newblock {\em Nat Cell Biol}, pages 1401--1410, 2008.

\bibitem{Lenne2021}
Pierre-Fran{\c{c}}ois Lenne, Jean-Fran{\c{c}}ois Rupprecht, and Virgile
  Viasnoff.
\newblock Cell junction mechanics beyond the bounds of adhesion and tension.
\newblock {\em Developmental Cell}, 56:202--212, 2021.

\bibitem{Chanet2017}
Soline Chanet, Callie~J. Miller, Eeshit~Dhaval Vaishnav, Bard Ermentrout,
  Lance~A. Davidson, and Adam~C. Martin.
\newblock Actomyosin meshwork mechanosensing enables tissue shape to orient
  cell force.
\newblock {\em Nature Communications}, 8:ncomms15014, 2017.

\bibitem{Ranft2010}
Jonas Ranft, Markus Basan, Jens Elgeti, Jean-Fran\c{c}ois Joanny, Jacques
  Prost, and Frank J\"ulicher.
\newblock Fluidization of tissues by cell division and apoptosis.
\newblock {\em Proc Natl Acad Sci U S A}, 107:20863--20868, 2010.

\bibitem{Zehnder2015}
Steven~M Zehnder, Marina~K Wiatt, Juan~M Uruena, Alison~C Dunn, W.~Gregory
  Sawyer, and Thomas~E Angelini.
\newblock Multicellular density fluctuations in epithelial monolayers.
\newblock {\em Phys Rev E Stat Nonlin Soft Matter Phys}, 92:032729, 2015.

\bibitem{Zehnder2015a}
Steven~M Zehnder, Melanie Suaris, Madisonclaire~M Bellaire, and Thomas~E
  Angelini.
\newblock Cell volume fluctuations in mdck monolayers.
\newblock {\em Biophys J}, 108:247--250, 2015.

\bibitem{Sui2018}
Liyuan Sui, Silvanus Alt, Martin Weigert, Natalie Dye, Suzanne Eaton, Florian
  Jug, Eugene~W. Myers, Frank Jülicher, Guillaume Salbreux, and Christian
  Dahmann.
\newblock Differential lateral and basal tension drive folding of drosophila
  wing discs through two distinct mechanisms.
\newblock {\em Nature Communications}, 9:4620, 2018.

\bibitem{Messal2019}
Hendrik~A. Messal, Silvanus Alt, Rute M.~M. Ferreira, Christopher Gribben,
  Victoria Min-Yi Wang, Corina~G. Cotoi, Guillaume Salbreux, and Axel Behrens.
\newblock Tissue curvature and apicobasal mechanical tension imbalance instruct
  cancer morphogenesis.
\newblock {\em Nature}, 566:126--130, 2019.

\bibitem{Wikipedia2022elastenglish}
\red{Wikipedia}.
\newblock Elastic moduli.
\newblock
  \href{https://en.wikipedia.org/wiki/Template:Elastic_moduli}{https://en.wikipedia.org/wiki/Template:Elastic\_moduli},
  2022.
\newblock Accessed: 2022-07-06.

\bibitem{landau1986theory}
L.D. Landau, E.M. Lifshitz, A.M. Kosevich, J.B. Sykes, L.P. Pitaevskii, and
  W.H. Reid.
\newblock {\em Theory of Elasticity: Volume 7}.
\newblock Course of theoretical physics. Elsevier Science, 1986.

\bibitem{Wkp2022elasticbulkmodulus}
\red{Wikipedia}.
\newblock Bulk modulus.
\newblock
  \href{https://en.wikipedia.org/wiki/Bulk_modulus}{https://en.wikipedia.org/wiki/Bulk\_modulus},
  2022.
\newblock Accessed: 2022-07-06.

\bibitem{Wkp2022elasticshearmodulus}
\red{Wikipedia}.
\newblock Shear modulus.
\newblock
  \href{https://en.wikipedia.org/wiki/Shear_modulus}{https://en.wikipedia.org/wiki/Shear\_modulus},
  2022.
\newblock Accessed: 2022-07-06.

\bibitem{Wkp2022hookeslaw}
\red{Wikipedia}.
\newblock Hooke's law.
\newblock
  \href{https://en.wikipedia.org/wiki/Hooke%27s_law}{https://en.wikipedia.org/wiki/Hooke's\_law},
  2022.
\newblock Accessed: 2022-07-06.

\bibitem{Wilson1972}
\red{T A Wilson}.
\newblock A continuum analysis of a two-dimensional mechanical model of the
  lung parenchyma.
\newblock {\em Journal of Applied Physiology}, 33(4):472--478, oct 1972.

\bibitem{Boal2012}
David Boal.
\newblock {\em Mechanics of the Cell}.
\newblock Cambridge University Press, 2012.

\bibitem{Ingber2014}
\red{Donald E Ingber}, Ning Wang, and Dimitrije Stamenovi{\'{c}}.
\newblock Tensegrity, cellular biophysics, and the mechanics of living systems.
\newblock {\em Reports on Progress in Physics}, 77(4):046603, apr 2014.

\bibitem{Wang2002}
\red{Ning Wang}, Iva~Marija Tolić-Nørrelykke, Jianxin Chen, Srboljub~M
  Mijailovich, James~P Butler, Jeffrey~J Fredberg, and Dimitrije Stamenović.
\newblock Cell prestress. i. stiffness and prestress are closely associated in
  adherent contractile cells.
\newblock {\em Am J Physiol Cell Physiol}, 282(3):C606--C616, Mar 2002.

\bibitem{Etienne2015}
Jocelyn Étienne, Jonathan Fouchard, Démosthène Mitrossilis, Nathalie Bufi,
  Pauline Durand-Smet, and Atef Asnacios.
\newblock Cells as liquid motors: {Mechanosensitivity} emerges from collective
  dynamics of actomyosin cortex.
\newblock {\em Proceedings of the National Academy of Sciences},
  112:2740--2745, 2015.

\bibitem{Bambardekar2014}
Kapil Bambardekar, Raphaël Clément, Olivier Blanc, Claire Chardès, and
  Pierre-François Lenne.
\newblock Direct laser manipulation reveals the mechanics of cell contacts in
  vivo.
\newblock {\em Proceedings of the National Academy of Sciences},
  112(5):1416--1421, 2015.

\bibitem{Kruyt2007}
N.P. Kruyt.
\newblock On the shear modulus of two-dimensional liquid foams: A theoretical
  study of the effect of geometrical disorder.
\newblock {\em J. Appl. Mech.}, 74:560--567, 2007.

\bibitem{Fouchard2020}
Jonathan Fouchard, Tom P.~J. Wyatt, Amsha Proag, Ana Lisica, Nargess
  Khalilgharibi, Pierre Recho, Magali Suzanne, Alexandre Kabla, and Guillaume
  Charras.
\newblock Curling of epithelial monolayers reveals coupling between active
  bending and tissue tension.
\newblock {\em Proceedings of the National Academy of Sciences},
  117:9377--9383, 2020.

\bibitem{Wyatt2019}
Tom P.~J. Wyatt, Jonathan Fouchard, Ana Lisica, Nargess Khalilgharibi, Buzz
  Baum, Pierre Recho, Alexandre~J. Kabla, and Guillaume~T. Charras.
\newblock Actomyosin controls planarity and folding of epithelia in response to
  compression.
\newblock {\em Nature Materials}, 2019.

\end{thebibliography}

\appendix

\newpage
\section*{Appendix}
\section{Calculations with  \maxima :}
\label{sec:approche:maxima}

Analytical calculations were performed in the following order with \maxima:
\begin{enumerate}
\item Definition of the variables : geometry and forces
\item Rheological equations
\item Force balance equations
\item Geometry-related equations
\item Calculation of the rest state
\item First-order expansion about the rest state
\item Resolution of the resulting system of equations
\item Monolayer Young modulus
\item Strain along the perpendicular direction and Poisson ratio
\item Expression of other moduli
\end{enumerate}
We include the script used to obtain
Eqs.~(\ref{eq:K:s0:g:k:L:H}-\ref{eq:nu:s0:g:k:L:H}).

\medskip

\begin{lstlisting}
kill ( all ) $
/* Definition of variables */
/* geometrical variables about rest state */
assume ( a0 >0);
a : a0 + da$
b : b0 + db$
z : z0 + dz$
p : p0 + dp$
thb : thb0 + dthb$
tha : tha0$
tha0 : %pi /2 $
V : V0 + dV ;
X : 2* b * cos ( thb ) $
X0 : subst ([ db =0 , dthb =0] , X );
Y : a - b * sin ( thb ) $
Y0 : subst ([ da =0 , db =0 , dthb =0] , Y );
eqV : V = X * Y * z$
/* macroscopic deformation variables */
eqEX : EX = (X - X0 )/ X0$
EY : (Y - Y0 )/ Y0$
/* macroscopic stress variables */
Sxx : 1/( X * Y )*(2* suu_a * a * z * cos(tha)^2
    +2*2* suu_b * b * z * cos(thb)^2 +2* suu_c * X * Y ) - p * z$
Syy : 1/( X * Y )*(2* suu_a * a * z * sin(tha)^2
    +2*2* suu_b * b * z * sin(thb)^2 +2* svv_c * X * Y ) - p * z$
Szz : 1/ V *(2* svv_a * a * z +2*2* svv_b * b * z ) - p$
/* microscopic deformation variables  */
ea : da / a0$
eb : db / b0$
ez : dz / z0$
/* lists of variables for Taylor expansions */
dvec0 : [ da=0, db=0, dz=0 , dV=0 , dp=0, dtha=0, dthb=0, EX=0];
dvec : [ da, db, dz, dV, dp, dthb, EX ];
start : [0, 0, 0, 0, 0, 0, 0];
order : [1, 1, 1, 1, 1, 1, 1];
/* */
/* Rheological equations */
/* microscopic rheological variables  */
suu_a : s0l + 2 * gstarl * ea + (kstarl - gstarl) * (ea + ez)$
suu_b : s0l + 2 * gstarl * eb + (kstarl - gstarl) * (eb + ez)$
suu_c : s0h + 2 * gstarh * EX + (kstarh - gstarh) * (EX + EY)$ 
svv_a : s0l + 2 * gstarl * ez + (kstarl - gstarl) * (ea + ez)$
svv_b : s0l + 2 * gstarl * ez + (kstarl - gstarl) * (eb + ez)$
svv_c : s0h + 2 * gstarh * EY + (kstarh - gstarh) * (EX + EY)$ 
/* incompressibility */
eqBulk : dV = 0 $ 
/* */
/* Force balance equations */
/* vertex balance */
eqForces : 2* suu_b * sin ( thb )+ suu_a * sin ( tha )=0 $
/* boundary conditions  */
eqSyy : Syy = 0 $
eqSzz : Szz = 0 $
/* */
/* Calculation of the rest state */
eqSyy0 : Syy0 = 0$
eqSzz0 : Szz0 = 0$
eqSxx0 : Sxx0 = 0$
Sxx0 : subst ( dvec0, ev ( Sxx, eval ));
Syy0 : subst ( dvec0, ev ( Syy, eval ));
Szz0 : subst ( dvec0, ev ( Szz, eval ));
eqV0 : subst ( dvec0, ev ( eqV ));
eqForces0 : subst ( dvec0, ev ( eqForces ));
thb0 : rhs ( solve ( eqForces0, thb0 )[1]);
V0 : rhs ( solve ( ev( eqV0 ), V0 )[1]);
p0 : rhs ( factor ( solve ( ev ( eqSzz0, eval ), p0 ))[1]);
z0 : rhs ( solve ( factor ( ev ( eqSxx0, eval )), z0 )[1]);
b0 : rhs ( solve ( factor ( ev ( eqSyy0, eval, eval )), b0 )[1]);
z0 : ev ( z0 );
V0 : ev ( V0, eval );
p0 : ev ( p0 );
/* */
/* First-order expansion about the rest state */
eqForces : factor ( taylor ( ev ( eqForces ), dvec, start, order ));
eqSyy : ev ( eqSyy, eval, eval ) ;
eqSyy : factor ( taylor ( eqSyy, dvec, start, order ));
eqSzz : ev ( eqSzz, eval, eval ) ;
eqSzz : factor ( taylor ( eqSzz, dvec, start, order ));
eqV : factor ( taylor ( ev ( eqV, eval, eval ), dvec, start, order ));
eqBulk : ev ( eqBulk, eval, eval ) ;
eqBulk : factor ( taylor ( eqBulk, dvec, start, order ));
eqEX : factor ( taylor ( ev ( eqEX, eval, eval ), dvec, start, order ));
Syst : [ eqForces, eqSyy, eqSzz, eqV, eqBulk, eqEX ] $
/* */
/* Resolution of the resulting system of equations */
[ eqda, eqdb, eqdz, eqdthb, eqdp, eqdV ]: 
    solve ( Syst, [ da, db, dz, dthb, dp, dV ])[1] $
da : factor ( rhs ( eqda ));
db : factor ( rhs ( eqdb ));
dz : factor ( rhs ( eqdz ));
dthb : factor ( rhs ( eqdthb ));
dp : factor ( rhs ( eqdp ));
dV : factor ( rhs ( eqdV ));
/* */
/* Monolayer Young modulus */
Sxx : ev ( Sxx, eval ) ;
EstarH3 : factor ( taylor ( Sxx, dvec, start, order ))/ EX ;
/* */
/* Strain along perpendicular direction and Poisson ratio */
Ey : ( factor ( taylor ( EY, dvec, start, order ))) $
Ey : factor ( ev ( Ey )) $
nustarH3 : factor ( - Ey / EX );
/* */
/* Expression of other moduli */
KstarH3 : factor ( EstarH3 /2/(1 - nustarH3 ));
GstarH3 : factor ( EstarH3 /2/(1+ nustarH3 ));
MstarH3 : factor ( EstarH3 /(1+ nustarH3 )/(1 - nustarH3 ));
lambdastarH3 : factor ( MstarH3 * nustarH3 );
/* Non-afinity */
naff: factor((ev(ea)-Ey)/EX);
naff: factor(subst(s0h=Psi*s0l,naff));
\end{lstlisting}

\section{The amplitudes of complex moduli
does not decrease with frequency}
\label{sec:frequency:dependence}

Let us consider a system made of a number of rheological elements
arranged in parallel or in series, assuming that 
each element is either a spring or a dashpot
or a fractional element. 
Fractional elements have moduli of the form $m(\omega) = c_\beta\,(i\omega)^\beta$,
with real prefactors $c_\beta$
and exponents $\beta$ between $0$ (spring limit)
and $1$ (dashpot limit). Here, $m$ can be any kind of modulus: shear, compression, etc.

The complex modulus $m(\omega)$ of each element thus has the two following properties:
\textit{(i)}, both the real part and the imaginary part of $m(\omega)$ are non-negative,
and \textit{(ii)}, the same is true
of its derivative $\mathrm{d}m/\mathrm{d}\omega$ with respect to angular frequency.
For a combination in parallel of such elements,
moduli add up and so do their derivatives,
thus properties \textit{(i)} and~\textit{(ii)} are conserved.
As a consequence, they also have the property that
\textit{(iii)}, the magnitude $|m(\omega)|$ does not decrease with frequency
(it is constant in the case of springs).%

Combinations in series involve adding up compliances $1/m(\omega)$.
These display similar properties, this time with  non-negative real parts
and {\textit{non-positive} imaginary parts,
\red{
and vice-versa for derivatives.
These properties
} 
are also conserved under summation.
\red{
They revert to the initial properties upon inversion
} 
and also imply property \textit{(iii)}.

Thus, multiple combinations, in parallel and in series,
of elements whose moduli display properties \textit{(i)} and~\textit{(ii)},
such as springs, dashpots and fractional elements,
also display these properties.

From the above considerations,
we find that the magnitude $|m(\omega)|$
of the complex modulus $m(\omega)$ of the system
does not decrease with frequency.

\section{2D hexagonal geometry}
\label{sec:2d}

In this case, the geometry remains as before, but with $Z = 0$. For clarity, we keep the former 2D notations for cortical tension $\so$,
and where necessary,
we use the typical cell height, which we note $h_0$
and assume constant and equal to $10\,\mu$m.

As in Fig.~\ref{fig:monolayer:top:view},
cell cortices are indexed by $a$ or $b$ depending on their orientations.
Cortices along axis $y$, labeled $a$,
are characterized by a length $\la=\lo+\delta \la$, 
a 1D tension $h_0\,\sigma_{a}$,
equivalent to a 2D stress $\sigma_a$
and a rheology
$\sigma_{a}=\sigma_{0}+\mu \frac{\delta \la}{\lo}$
with generalized modulus $\mu$.
Similarly,  $b$ cortices  are characterized by $\lb=\lo+\delta \lb$,
and $\sigma_{b}=\sigma_{0}+\mu \frac{\delta \lb}{\lo}$,
as well by their
orientation $\theta=\frac{\pi}{6}+\delta\theta$.
Force balance at a vertex now reads
$2\sigma_{a} -4 \, \sigma_{b} \sin(\theta)=0$
where, by symmetry, cortices $a$ remain along direction $y$
upon tissue traction along the $x$-axis,
(see Eq.~\eqref{eq:force:balance:six:cortex:tensions} for comparison).
The macroscopic (2D) stress components are expressed as a function of
microscopic parameters as:
\begin{eqnarray}
  \Sigma_{xx} &=& \frac{h_0}{S} \, 4 \lb \sigma_{b} \cos^{2}\theta - \pdd \,,  
  \label{eq:stress:macro:H2:xx}
  \\
\Sigma_{yy}&=&  \frac{h_0}{S} \left( 2\la\, \sigma_{a}
+  4 \lb\, \sigma_{b}\,\sin^2\theta \right) - \pdd   \,,
               \label{eq:stress:macro:H2:yy}
\end{eqnarray}
(see Eqs.~(\ref{eq:stress:macro:xx}-\ref{eq:stress:macro:yy}) for comparison).

\subsection{Incompressible case}

We first consider the intra-cellular material to behave as an inviscid,
incompressible fluid, with pressure $\pdd$ and constant \textit{area} 
$S = \LX\LY = S_0$. 

Our calculation is performed 
along the same lines as in the 3D case.
We find a diverging tissue in-plane bulk modulus $\Ktis_\mathrm{2D}$ 
(which reflects the assumed incompressibility
of each cell  within the tissue, $S=S_0$)
while the shear modulus reads:
\begin{equation}
\Gtis_\mathrm{2D}(\omega) = \frac{\sigma_0}{
\sqrt{3}}\,
\,\frac{\sigma_0+3\mu(\omega)}{\sigma_0+\mu(\omega)}  \,.
\label{eq:H2:G:brut}
\end{equation}
The Young and Poisson moduli are obtained according to 
Eqs.~(\ref{eq:E:s0:g:k:L:H}-\ref{eq:nu:s0:g:k:L:H}):
since $\Ktis_\mathrm{2D} = \infty$,
$E_\mathrm{2D}= 4 \, \Gtis_\mathrm{2D}$ and 
$\nutis_\mathrm{2D}= 1$.
In the high-frequency limit $\omega \to \infty$,
where $|\mu(\omega)|$ becomes much larger than $\sigma_0$,
expression~\eqref{eq:H2:G:brut} is 
bounded and yields
$ E_\mathrm{2D}(\infty) =
4 \sqrt{3} \sigma_0$,
which is independent of frequency and real, \textit{i.e.,} 
it corresponds to an elastic rheology,
at variance with the power-law behaviour observed experimentally
in this limit \cite{Bonfanti2020}.
In addition, for a realistic value of the cortical tension 
$\so = 0.3$~mN/m~\cite{Salbreux2012}, 
we obtain the high frequency limit $E_\mathrm{2D}(\infty) = 2.1$~mN/m,
short of the order of magnitude observed experimentally in suspended monolayers 
($E_\mathrm{2D} \sim 200$ mN/m, corresponding to 
$E_\mathrm{3D} = 20 \pm 2$ kPa measured in \cite{Harris2013} for $h_0 = 10\,\mu$m).

\subsection{Compressible case}

In an attempt to circumvent this unsuitable elastic behaviour 
of our 2D model in the high frequency limit,
we questioned
the cell incompressibility assumption
in the 2D calculation. 
Indeed, the 3D cell \textit{volume} conservation
expected on the time scales relevant to this study is compatible with 
in-plane variations of the apical cell surface, 
as the cell height varies correspondingly
so as to conserve cell volume.
We tested this hypothesis by replacing the incompressibility condition $S = S_0$ 
by a condition on apical surface variations
$\mathrm{d}S/S_{0}=- \mathrm{d}\pdd / \Kcytodd$
involving the cell pressure $\pdd$ and a 2D cell compression modulus $\Kcytodd$. 
In this case, we find the following expressions of macroscopic
moduli which generalize the previous results
to finite values of the cell compression modulus $\Kcytodd$:
\begin{eqnarray}
  \label{eq:H2c:K:brut}
\Ktis_\mathrm{2Dc} &=&
\left(\Kcytodd-\frac{\so}{\sqrt{3}}\right)
+\frac{\Mucortex}{\sqrt{3}} \,,
  \\
  \label{eq:H2c:G:brut}
\Gtis_\mathrm{2Dc} &=&
\frac{\so}
{\sqrt{3}}\,
\frac{\so+3\Mucortex}
{\so+\Mucortex} \,,
  \\
\frac{1}{\Etis_\mathrm{2Dc}} &=& \frac{1}{4\,\Ktis_\mathrm{2Dc}} 
+ \frac{1}{4\,\Gtis_\mathrm{2Dc}} \,,
  \label{eq:H2c:E:brut}
\\
  \label{eq:H2c:nu:brut}
  \nutis_\mathrm{2Dc}
 &=&
1-\frac{2\,\so\,(\so +3\Mucortex)
}
{
\sqrt{3}\,\Kcytodd (\so+\Mucortex)
+\Mucortex(3\so+\Mucortex)
} \,.
\end{eqnarray}

According to Eq.~\eqref{eq:H2c:E:brut}, the high-frequency $\omega \to \infty$
behaviour of the macroscopic Young's modulus is however unchanged, 
with an elastic limit $E_\mathrm{2Dc}(\omega) \to 4 \sqrt{3}\,\so$,
and remains incompatible with the power-law behaviour observed in this limit.
It is easy to show that the complex values of $\Ktis_\mathrm{2Dc}$
and $\Gtis_\mathrm{2Dc}$ belong to the same quadrant of the complex plane.
As a consequence, the modulus of $\frac{1}{\Etis_\mathrm{2Dc}}$ is always
larger than the modulus of  $\frac{1}{4 \Gtis_\mathrm{2Dc}}$.
This implies that the modulus of $\Etis_\mathrm{2Dc}$ is always smaller than
its high-frequency limit $ E_\mathrm{2Dc}(\infty) = 4 \sqrt{3} \sigma_0$.
As was already the case for an incompressible 2D tissue, the 
order of magnitude of the Young's modulus obtained in the compressible case 
is incompatible with the experimentally measured value.

\subsection{Disordered 2D monolayer}

Simulations will be needed to obtain the corresponding results
for disordered 2D monolayers,
with cells differing for instance in surface area, 
edge length~\cite{Kruyt2007},
number of first neighbours or cortex tensions or moduli.
Any kind of disorder will in general lead to unequal cell pressures
and, correspondingly, to edge curvatures.
However, two limiting behaviours can be anticipated.

\textit{(i)} At low frequencies, since the cortex modulus
is negligible as compared to the rest tension $\so$,
the disordered monolayer will behave elastically at small deformations.
Its moduli will be proportional to $\so$ (just like in the ordered case)
as results from dimensional analysis, 
a fact that is well known in the liquid foam community~\cite{Weaire01,Cantat2013}.

\textit{(ii)} By contrast, in the large frequency limit,
unless the cortex rheology is purely elastic in that limit,
the cortex modulus will dominate over the rest tension
($|\mu|/\so\to\infty$).
Thus, unless all cortices remain undeformed at first order,
the corresponding forces proportional to $\mu$
will become dominant within the macroscopic stress.
Now, as for cortex deformation, it is to be expected that,
for general disordered networks as opposed to the ordered, honeycomb structure,
there will exist no deformation mode
that will conserve both each cell cytoplasm surface area
\textit{and} each edge length, at first order.
As a result, any macroscopic deformation
will result in at least some cell-cell junctions changing their lengths.
The corresponding tensions, and hence the macroscopic stress,
will therefore be sensitive to the cell cortex rheological modulus, $\mu$.

\section{\red{Affine behaviour}}
\label{sec:affinity}

\red{As mentioned in Section~\ref{sec:resolution},
some elements in the cell deform in a non-affine manner,
that is, not in the same proportions as the monolayer.}

\red{In order to demonstrate that, let us consider
how two quantities are affected by the applied deformation $\epsx$, namely:
\textit{(i)} the deformation $\epsa=\la/\lo$ of the $a$ facets,
defined by Eq.~\eqref{eq:epsa},
and \textit{(ii)} the transverse deformation $\epsy$ 
of the representative volume, defined by Eq.~\eqref{eq:epsy}.
Of course, because we are considering linear response,
they are both proportional to the applied deformation $\epsx$.}
\red{For the sake of clarity, let us assume that the monolayer 
is being stretched in the $x$-direction ($\epsx>0$).
As a result, it shrinks in the $y$-direction ($\epsy<0$).}
\red{As for the $a$ facets, three options are possible,
as shown on Fig.~\ref{fig:def:modes}.
The $a$ facets can shrink exactly like the sample (affine deformation  $\epsa=\epsy<0$, Fig.~\ref{fig:def:modes}b).
They can shrink more than the sample ($\epsa<\epsy<0$, Fig.~\ref{fig:def:modes}c).
They can shrink less ($\epsy<\epsa\leq 0$, Fig.~\ref{fig:def:modes}d).
} 

\red{Hence, it appears reasonable to evaluate
the difference between $\epsa$ and $\epsy$,
divided by the applied deformation,
and choose it as an \textit{indicator of non-affinity}
for the monolayer subjected to the present deformation,
which we name $\nonaffinity$:}
\begin{eqnarray}
\red{\nonaffinity}&\red{\equiv}&
\red{
\frac{\epsa-\epsy}{\epsx}
}
\\
&\red{=}&\red{
\frac{(\kl+\gl-\sol)
(2\kh+\Psi(3\sol+\kl+9\gl))}
{\mathcal{D}} \,,
}
\\
\red{\mathcal{D}} &\red{=}& \red{
4\Psi\sol^2+2(\kh+\gh)\sol
+(7\kl+15\gl)\Psi\sol  \,,
}
\nonumber\\
&&\red{
+2(\kh+\gh)(\kl+\gl)
+\Psi(\kl^2+10\gl\kl+9\gl^2)  \,.
}
\end{eqnarray}
\red{The non-affinity indicator $\nonaffinity$
turns out to be nonzero unless a very specific condition is met:}
\begin{equation}
\red{\kl+\gl=\sol  \,.}
\label{eq:condition:affinity}
\end{equation}
\red{This expression can be partly understood qualitatively.}
\red{In the limit where the facet moduli $\kl$ and $\gl$ 
are much smaller than the rest tension $\sol$
(this corresponds to the low frequency limit),
the tensions will remain unchanged by the deformation,
and it is expected that the angles between lateral facets,
which result from the force balance,
remain equal to their initial value, $2\pi/3$
(see Fig.~\ref{fig:def:modes}c).
It is obvious that the $a$ facets then shrink more
than the overall sample (see blue lines for comparison).
And indeed, $\nonaffinity$ is then negative,
consistently with the fact that $\kl,\,\gl\ll\sol$.
}
\red{By contrast, when $\kl$ and $\gl$ 
are much larger than $\sol$
(this corresponds to the high frequency limit),
the facets are purely elastic and their dimensions tend to keep constant dimensions
(see Fig.~\ref{fig:def:modes}d).
It is obvious that the $a$ facets then shrink less
than the overall sample, and indeed, $\nonaffinity$ is positive in this limit.
} 

\begin{figure}[t]
\centerline{\includegraphics[scale=1.0]{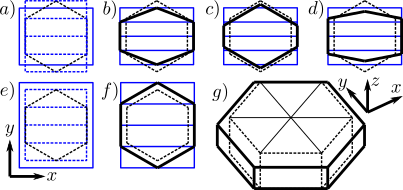}}
\caption{
\red{
\textbf{Deformation modes. ($a$-$d$)}:
the monolayer is stretched in the $x$-direction.
\textbf{($e$-$g$)}: 
the monolayer is stretched isotropically in the $x$-$y$ plane.
\textbf{($a$, $e$)}: the blue dashed rectangle with intermediate lines is a guide for the eye
that matches the undeformed (dashed) hexagon.
The corresponding rectangle after deformation
is drawn with a blue, solid line.
\textbf{($b$, $f$)}: the deformation is affine 
(hexagon deformation matches overall deformation, see intermediate blue lines,
which implies $\epsa=\epsy$),
when \textbf{($b$)} $\nonaffinity=0$ 
or when \textbf{(f)} the deformation is isotropic.
\textbf{($c$)}: 
$a$ facets shrink more ($\epsa<\epsy<0$, $\nonaffinity<0$) 
than the overall sample (compare deformed hexagon and intermediate blue lines).
\textbf{($d$)}: 
$a$ facets shrink less ($\epsy<\epsa\leq 0$, $\nonaffinity>0$)
than the overall sample (compare deformed hexagon and intermediate blue lines).
\textbf{($g$):} under isotropic in-plane stretching,
while horizontal facets are stretched isotropically,
lateral facets are elongated along their respective horizontal directions in the $x$-$y$ plane
and shrink along direction $z$.
}
}
\label{fig:def:modes}
\end{figure}

\red{
Let us recall that the calculation in the present work
was carried out 
using the simplifying assumption, stated in Section~\ref{sec:resolution},
that basal and apical facets deform affinely, \textit{i.e.}, according to the overall deformation, measured by $\epsx$ and $\epsy$.
That assumption is expressed through Eq.~\eqref{eq:rheo:cortex:c:xx} for the stress in those facets (assumed uniform),
as well as through Eq.~\eqref{eq:force:balance:six:cortex:tensions}
which states that the vertical edge where lateral facets meet
receives no forces from the horizontal facets.
} 

\red{
Whenever $\epsa$ and $\epsy$ differ according to the present calculation,
the edges where lateral facets meet (hexagon corners) and the initially corresponding points of the horizontal facets
\textit{do not coincide} after the overall deformation has been applied
(see mismatch between the hexagon corners and the intersections of blue lines in Figs.~\ref{fig:def:modes}c-d).
Yet such a point is in reality the corner of three neighbouring cells, and obviously the horizontal facets
should remain attached to the lateral facets!
} 
\red{
Thus, in reality these corresponding points are maintained attached to each other (through a tensile force)
and the real deformation of facets $a$
adopts some intermediate value between $\epsy$ and the value of $\epsa$ calculated here.
In other words, the sign of $\nonaffinity$ is correct,
but its magnitude is somewhat overestimated
in the framework of the present assumption.
} 

\red{
Note that unless the cortex rheology is purely elastic
in some frequency range (\textit{i.e.}, with real moduli),
there is no frequency where condition~\eqref{eq:condition:affinity} can be satisfied,
hence the deformation is always somewhat non-affine.
} 

\red{
For the sake of completeness, let us mention
that we have also explored several refined geometrical descriptions
of the horizontal facet kinematics.
Although they did fix the mismatch between lateral facet vertical edges and horizontal facets,
none of them was both reasonably simple and six-fold symmetric.
Hence, we resolved to keep the present version,
albeit somewhat unconsistent as detailed in the present Appendix.
Only with a highly refined mesh both in the horizontal and the lateral facets can one hope to fully deal with this issue
and calculate the exact amplitude of the non-affine deformations in the monolayer.
That is beyond the scope of the present work.
} 

\end{document}